\documentclass[journal]{IEEEtran}
\usepackage{latexsym}
\usepackage{amsfonts,amssymb,amsmath}
\usepackage{hyperref}
\def\BibTeX{{\rm B\kern-.05em{\sc i\kern-.025em b}\kern-.08em T\kern-.1667em\lower.7ex\hbox{E}\kern-.125emX}}
\usepackage{algorithm}
\usepackage{algpseudocode}
\usepackage{amsmath}
\usepackage{amsfonts}
\usepackage{graphicx}
\usepackage{epsfig}
\usepackage{cite}
\usepackage{txfonts}
\usepackage{relsize}
\usepackage{color}
\usepackage{color}
\usepackage{CJK}
\usepackage{times}
\usepackage{multicol}
\usepackage{stfloats}
\usepackage{float}
\usepackage{lipsum}
\usepackage{bm}
\usepackage{subfigure}

\begin{document}

\title{STAR-RIS-Assisted Cell-Free Massive MIMO with Multi-antenna Users and Hardware Impairments Over Correlated Rayleigh Fading Channels
\thanks{This work was supported by the Research Grants Council under the Area of Excellence scheme grant AoE/E-601/22-R.}}
\author{Jun~Qian,~\IEEEmembership{Member,~IEEE,}
Ross~Murch,~\IEEEmembership{Fellow,~IEEE}~and~Khaled~B.~Letaief,~\IEEEmembership{Fellow,~IEEE}
   
\thanks{This work was supported by the Hong Kong Research Grants Council with Area of Excellence grant AoE/E-601/22-R.}}

\author{Jun~Qian,~\IEEEmembership{Member,~IEEE,}
        Ross~Murch,~\IEEEmembership{Fellow,~IEEE,}
and~Khaled~B.~Letaief,~\IEEEmembership{Fellow,~IEEE}
\thanks{The authors are with the Department of Electronic and Computer Engineering, The Hong Kong University of Science and Technology,
Hong Kong (e-mail: eejunqian@ust.hk, eermurch@ust.hk, eekhaled@ust.hk).}}

\maketitle
{\begin{abstract}

Integrating cell-free massive multiple-input multiple-output (MIMO) with simultaneous transmitting and reflecting reconfigurable intelligent surfaces (STAR-RISs) can provide ubiquitous connectivity and enhance coverage.
This paper explores a STAR-RIS-assisted cell-free massive MIMO system featuring multi-antenna users, multi-antenna access points (APs), and multi-element STAR-RISs, accounting for transceiver hardware impairments. We first establish the system model of STAR-RIS-assisted cell-free massive MIMO systems with multi-antenna users. Subsequently, we analyze two uplink implementations: local processing and centralized decoding (Level 1), and fully centralized processing (Level 2), both implementations incorporating hardware impairments. We study the local and global minimum mean square error (MMSE) combining schemes to maximize the uplink spectral efficiency (SE) for Level 1 and Level 2, respectively. The MMSE-based successive interference cancellation detector is utilized to compute the uplink SE. We introduce the optimal large-scale fading decoding at the central processing unit and derive closed-form SE expressions utilizing maximum ratio combining at APs for Level 1. Our numerical results reveal that hardware impairments negatively affect SE performance, particularly at the user end. However, this degradation can be mitigated by increasing the number of user antennas. Enhancing the number of APs and STAR-RIS elements also improves performance and mitigates performance degradation. Notably, unlike conventional results based on direct links, our findings show that Level 2 consistently outperforms Level 1 with arbitrary combining schemes for the proposed STAR-RIS-assisted system.


\end{abstract} 

}

\maketitle

\section{Introduction}

\IEEEPARstart{M}{assive} multiple-input multiple-output (MIMO) has been a key technology in successfully meeting the increasing demand for connectivity and higher data rates in wireless networks\cite{9570143,10167480,8388873}. However, despite its advantages, strong inter-cell interference remains a challenge for cell-boundary users in cellular networks\cite{10163977,9665300}. Consequently, new technologies are required to address inter-cell interference and support massive MIMO communication structures. These potential technologies are vital to achieving ubiquitous and robust connectivity in future wireless networks \cite{10201892,10163977}.

Cell-free massive MIMO has been deemed an appealing technology for future wireless communication, eliminating inter-cell interference and enabling an amorphous network \cite{9737367,10163977,10058895,7827017}. The fundamental structure of cell-free massive MIMO is based on integrating massive MIMO and distributed networks comprising numerous geographically distributed access points (APs) connected to a central processing unit (CPU) via fronthaul links. In this configuration, all APs cooperate with the CPU, serving all users simultaneously without cell boundaries\cite{9416909,7827017,qian2024performance}. In \cite{8845768}, the authors outlined four implementations of cell-free massive MIMO, ranging from fully centralized processing to fully distributed processing, utilizing global or local minimum mean-square error (MMSE) combining to maximize the spectral efficiency (SE) performance.
Moreover, \cite{8845768} indicated that the centralized implementation with MMSE processing could make the cell-free massive MIMO competitive. Most existing works rely on an effective two-layer decoding scheme that engages an arbitrary combining scheme at the APs, followed by a large-scale fading decoding (LSFD) method at the CPU \cite{9737367,8845768,10201892,10163977,qian2024performance}. 
Implementing advanced processing schemes in cell-free massive MIMO can significantly boost SE performance \cite{10297571}. However, increasing APs results in higher power consumption and network overhead and experiences poor quality of service (QoS) under adverse propagation conditions \cite{9665300,10167480,10058895}.

To address the limitations of cell-free massive MIMO systems, reconfigurable intelligent surfaces (RISs) with passive reflective elements have emerged as an appealing technology to assist cell-free massive MIMO systems \cite{9322151,10167480}. RIS elements can be adjusted to shape electromagnetic waves and introduce additional reconfigurable cascaded links, yielding passive beamforming and effective energy savings \cite{9326394,10058895,10297571}. Integrating RIS with cell-free massive MIMO systems has demonstrated substantial potential in improving SE performance under harsh propagation conditions\cite{10225319,9322151,10167480}. Given that RISs are configured to reflect incoming signals, the receiver and transmitter are therefore located on the same side of RISs \cite{10316600,10297571}. However, in real-world scenarios, users may be positioned on both RIS sides \cite{9570143,10297571}. To address coverage limitations, \cite{9437234,9690478} introduced the innovative concept of simultaneously transmitting and reflecting RIS (STAR-RIS), facilitating full space coverage. STAR-RIS can manage the separation of incoming signals into reflected and transmitted components by controlling phase shifts and amplitudes\cite{9570143,10297571,10264149,qian2024performance}. 
Inspired by integrating RISs and cell-free massive MIMO systems, the interplay between STAR-RISs and cell-free massive MIMO systems has also attracted research interest. However, this interplay is still at an early stage \cite{10297571}.
The SE analysis of spatially correlated STAR-RIS-assisted and active STAR-RIS-assisted cell-free massive MIMO was introduced in \cite{10297571} and \cite{10264149}, respectively. Then, \cite{10316600} optimized the sum rate of STAR-RIS-assisted cell-free massive MIMO systems. \cite{qian2024performance} designed a projected gradient descent algorithm to investigate and boost the SE performance of STAR-RIS-assisted cell-free massive MIMO systems experiencing electromagnetic interference and phase errors. 

Most current studies on STAR-RIS-assisted cell-free massive MIMO have considered ideal/high-quality hardware conditions at both transmitters and receivers\cite{qian2024performance,10316600,10297571,10264149}. However, practical transceivers often incorporate
non-ideal hardware components, which can introduce hardware impairments, such as carrier-frequency offset, nonlinearities of analogue components and oscillator phase noise \cite{10418910,10225319,8476516}. In this case, hardware impairment distorts the transmit and receive signals to cause channel estimation errors and severe
performance loss \cite{10418910,10225319,10201892,9528977}. Thus, it is vital to investigate system performance affected by hardware impairments. Research on cell-free massive MIMO systems incorporating hardware impairments has garnered attention \cite{10445267,8476516,10163977}.
In \cite{8476516}, the authors introduced the generic model of hardware impairments into 
cell-free massive MIMO systems and proposed a max-min power control algorithm to enhance performance. \cite{10445267} investigated the total energy efficiency of wireless-powered, hardware-impaired, cell-free massive MIMO systems. Meanwhile, \cite{10163977} explored how hardware impairments affect user-centric cell-free massive MIMO systems with multi-antenna users. In the context of deploying RISs and STAR-RISs, \cite{10418910} assessed the RIS-assisted system performance with the joint effects of channel aging and hardware impairments. \cite{10225319} studied the uncorrelated RIS-assisted cell-free massive MIMO system affected by hardware impairments, proposing a modified channel estimation scheme. \cite{10475146} introduced a robust beamforming method for STAR-RIS combined with the active base-station beamforming to maximise the hardware-impaired sum rate. \cite{10264820} derived the minimum signal-to-interference-plus-noise-ratio (SINR) achieved by the optimal linear precoder in the STAR-RIS-assisted massive MIMO with hardware impairments. These observations underscore the necessity of considering hardware impairments in practical deployment scenarios. However, as far as we know, only authors of \cite{sui2024starrisaidedcellfreemassivemimo,10841966} addressed the downlink performance of the STAR-RIS-assisted cell-free massive MIMO involving hardware impairments. The scarcity of studies on hardware impairments in STAR-RIS-assisted cell-free massive MIMO systems underlines the rationale for our work.

The majority of existing studies have focused on cell-free massive MIMO systems with single-antenna users \cite{9416909,8845768}, similar to RIS/STAR-RIS-assisted cell-free massive MIMO systems \cite{10326460,9322151,10167480,sui2024starrisaidedcellfreemassivemimo,10841966}. However, as antenna technology advances, contemporary users/devices are expected to utilize multiple antennas to improve multiplexing gain and system performance. This has been shown as efficient and effective in cell-free massive MIMO systems \cite{10201892,10163977,9737367,10571171,9079911}.  The authors in \cite{9737367} explored the uplink performance of cell-free massive MIMO incorporating multi-antenna users, demonstrating significant benefits of introducing additional user antennas. \cite{9079911} introduced downlink pilots on the downlink performance of cell-free massive MIMO systems with multi-antenna users. \cite{10571171} studied the joint impact of spatial correlation and mutual coupling of cell-free massive MIMO with multi-antenna users. Moreover, \cite{10201892,10163977} explored the effect of hardware impairments on cell-free massive MIMO with multi-antenna users, considering superimposed pilots and user-centric structure, respectively. Additionally, the role of multi-antenna users has been documented in optimizing designs for RIS-assisted cell-free massive MIMO systems. \cite{10058895} examined the design of active and passive beamforming for uncorrelated RIS-aided cell-free massive MIMO systems with multi-antenna users. \cite{10620555} introduced a joint precoding algorithm for a multi-RIS-aided cell-free massive MIMO network with multi-antenna users to maximize the weighted sum rate. However, the analysis of jointly spatially correlated channels in RIS-assisted networks with multi-antenna users remains significantly absent, as does the consideration of multi-antenna users in STAR-RIS-assisted configurations. This includes the absence of an analysis of design guidelines and performance limits. Given
these observations, there is a vital need to conduct a performance analysis of STAR-RIS-assisted cell-free massive MIMO systems with multi-antenna users in the context of jointly spatially correlated channels.

To the best of our knowledge, the impact of hardware impairments on jointly spatially correlated MIMO channels in STAR-RIS-assisted cell-free massive MIMO systems with multi-antenna users has not been previously studied. This paper is the first to conduct a performance analysis of jointly spatially correlated STAR-RIS-assisted cell-free massive MIMO systems featuring multi-antenna users, where both APs and users suffer from non-ideal hardware impairments. We specifically analyze two uplink implementations: local processing and centralized decoding (Level 1) and fully centralized processing (Level 2), as outlined in \cite{8845768}. We present local and global MMSE combining schemes for respective implementations and propose the optimal LSFD scheme for Level 1. The primary contributions are delineated as follows:
\begin{itemize}
\item We establish a novel STAR-RIS-assisted cell-free massive MIMO model that incorporates multi-antenna users, marking a significant advancement in the field. Our model features jointly spatially correlated Rayleigh fading channels, accounting for the realistic scenario where all APs and users are non-ideal and subject to hardware impairments. To the best of our knowledge, this is the first study to integrate both multi-antenna users and hardware impairments into STAR-RIS-assisted cell-free massive MIMO systems, thereby contributing new insights into their performance and applicability.

\item We analyze the uplink SE with hardware impairments for two uplink implementations based on the MMSE-based successive interference cancellation (MMSE-SIC) detector. The MMSE-SIC detector serves as an optimal receiver architecture for parallel data streams \cite{10163977}. We explore local and global MMSE combining schemes that maximize SE for respective implementations. Additionally, the optimal LSFD decoding scheme for Level 1 is introduced. We derive the closed-form SE expressions for Level 1 utilizing maximum ratio (MR) combining.

\item Monte Carlo (MC) simulations verify the closed-form analytical results examining system performance. The results reveal that multi-antenna users can offset the performance degradation caused by hardware impairments. Moreover, the introduction of more APs, antennas per AP, and STAR-RIS elements can enhance system performance. Notably, Level 2 can significantly improve SE for the proposed STAR-RIS-assisted system. 

\end{itemize}

The remainder of this paper is outlined as follows. Section II models the spatially correlated STAR-RIS-assisted channel model with multi-antenna users. Section III introduces the uplink channel estimation with pilot contamination and hardware impairments. We derive the uplink SE with hardware impairments for two uplink implementations in Section IV and provide numerical results in Section V. Finally, Section VI concludes the paper and proposes avenues for future work.

\section{System Model}
\begin{figure}[!t]
\centering
\includegraphics[width=0.85\columnwidth]{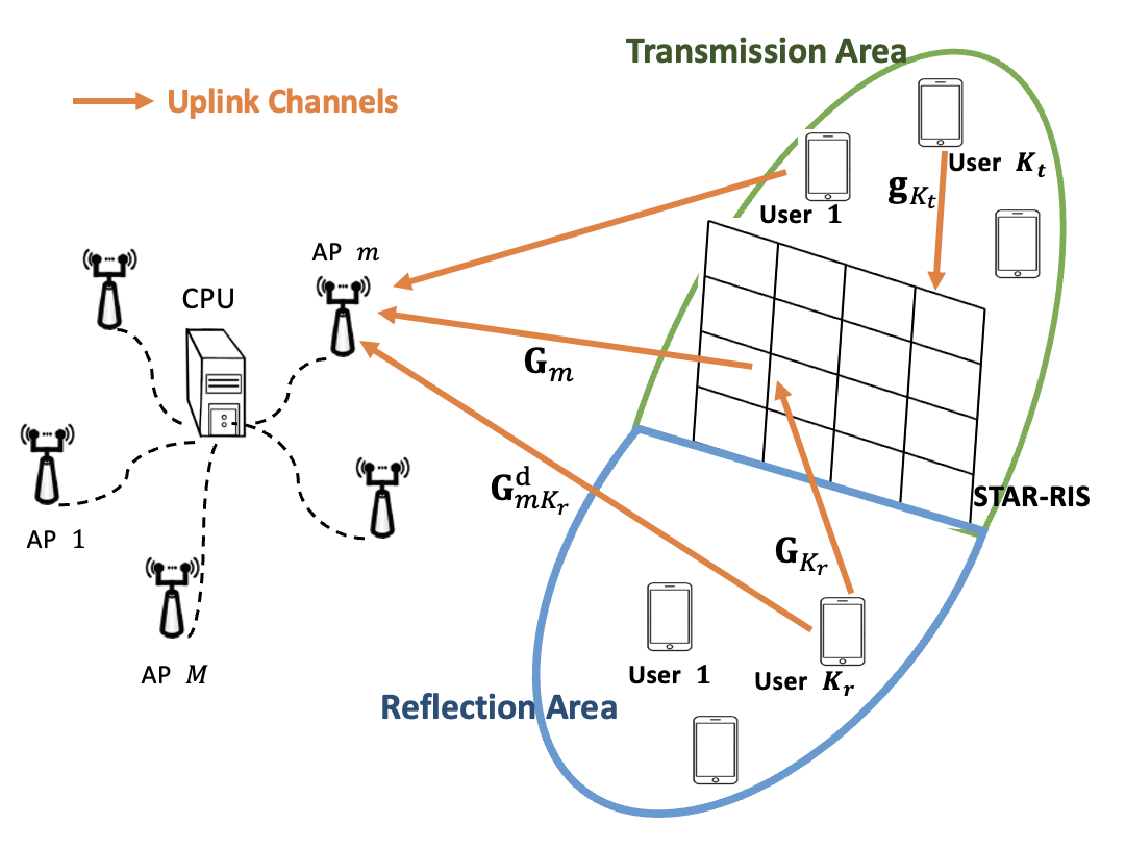}
\caption{A STAR-RIS-Assisted Cell-Free Massive MIMO System deployment.}
\label{fig_11}
\end{figure}
As depicted in Fig. \ref{fig_11}, this paper investigates a STAR-RIS-assisted cell-free massive MIMO system\cite{10297571,10264149}. $M$ APs, each with $N_{ap}$ antennas and optimally connected to the CPU via fronthaul links, can coherently serve $K$ users, each with $N_{u}$ antennas.
The communication between the APs and the users is facilitated by an $L$-element STAR-RIS. We define two sets of users: $\mathcal{K}_r$ and $\mathcal{K}_t$ with the respective set cardinality $|\mathcal{K}_r|$=$K_r$ and $|\mathcal{K}_t|$=$K_t$ ($K_r+K_t=K$, $\mathcal{K}_r\cap\mathcal{K}_t=\varnothing$). Users in $\mathcal{K}_r$ are positioned in the reflection area (the same side of the APs), while those in $\mathcal{K}_t$ are in the transmission area.  
The STAR-RIS operation mode for the $k$-th user is represented as $\omega_k,~\forall k$. Specifically, $\omega_k=r$ if the $k$-th user is located in the reflection area with the STAR-RIS reflection mode ($k \in \mathcal{K}_r$); otherwise, $\omega_k=t$ when the $k$-th user resides in the transmission area with the STAR-RIS transmission mode ($k \in \mathcal{K}_t$). We also define $\mathcal{W}_k$ as the set of users that share the same STAR-RIS operation mode, including the $k$-th user \cite{10297571}. 

\subsection{STAR-RIS Protocol}

STAR-RIS operation usually follows three feasible protocols: energy splitting (ES), mode switching (MS), and time switching (TS)\cite{9570143,10297571}. This work focuses primarily on the ES protocol, where all STAR-RIS elements function jointly in reflection and transmission modes to serve all users, irrespective of their locations. The STAR-RIS coefficient matrices are
$\boldsymbol{\Theta}_{t}=\text{diag}(u_{1}^t{\theta}_{1}^t,u_{2}^t{\theta}_{2}^t,...,u_{L}^t{\theta}_{L}^t)\in\mathbb{C}^{L\times L}$ for the transmission mode and $\boldsymbol{\Theta}_{r}=\text{diag}(u_{1}^r{\theta}_{1}^r,u_{2}^r{\theta}_{2}^r,...,u_{L}^r{\theta}_{L}^r)\in\mathbb{C}^{L\times L}$ for the reflection mode. In this context, the induced phase shifts are ${\theta}_{l}^t=e^{i\varphi_{l}^t},~{\theta}_{l}^r=e^{i\varphi_{l}^r}$ with $\varphi_{l}^t,~\varphi_{l}^r \in[0,2\pi),~\forall l$. The ES amplitude coefficients are $u_{l}^t,~u_{l}^r \in[0,1]$ with $(u_{l}^t)^2+(u_{l}^r)^2=1$\cite{9570143,10264149}. Note that the coupling between the reflecting and transmitting coefficient phases is intended for future work \cite{10297571}. 
Moreover, we can find that the MS protocol is a particular instance of the ES protocol, where $u_{l}^t,~u_{l}^r$ are restricted to binary values, that is, $u_{l}^t,~u_{l}^r \in\{0,1\}$
\cite{9570143,10297571}.

\subsection{STAR-RIS-Assisted Channel Model}
In this work, the system operates in time-division duplex (TDD) mode, facilitating channel reciprocity \cite{9905943,10571171}. Since multi-antenna APs and users are deployed, the spatial correlations associated with the transmitter and receiver cannot be overlooked \cite{10163977,9875036,9905943}. To this end, the jointly spatially-correlated Rayleigh channel model is characterized by the conventional Kronecker model\cite{9905943,10163977,7500452,9685245}. The aggregate uplink channel from the $k$-th user to the $m$-th AP, $\textbf{G}_{mk}\in \mathbb{C}^{N_{ap}\times N_{u}}$, is given by
\begin{equation}
     \displaystyle \textbf{G}_{mk}=\textbf{G}_{mk}^{\text{d}}+\textbf{G}_{mk}^{\text{c}},
     \label{aggregate_uplink_channel}
   \end{equation}
where $\textbf{G}_{mk}^{\text{d}}\in \mathbb{C}^{N_{ap}\times N_{u}}$ is the direct channel from the $k$-th user to the $m$-th AP, given by
\begin{equation}
     \displaystyle \textbf{G}_{mk}^{\text{d}}=\sqrt{\beta_{mk}}{\textbf{R}}_{m,r}^{1/2}\textbf{v}_{mk}{\textbf{R}}_{k,t}^{1/2},
     \label{direct_uplink_channel}
   \end{equation}
where $\beta_{mk}$ is the large-scale fading coefficient between the $m$-th AP and the $k$-th user, $\textbf{v}_{mk}\in\mathbb{C}^{N_{ap}\times N_u}$ refers to the independent fast-fading channel composed of independent and identically
distributed (i.i.d) random variables, distributed as $\mathcal{CN}(0,1)$ \cite{qian2024performance}. The spatial correlation matrix for the $m$-th AP is denoted as ${\textbf{R}}_{m,r} \in \mathbb{C}^{N_{ap}\times N_{ap}}$. The spatial correlation matrix at the $k$-th user is represented by ${\textbf{R}}_{k,t} \in \mathbb{C}^{N_u\times N_u}$, following Jake's model, with $[{\textbf{R}}_{k,t}]_{n,m}=J_0\left(2\pi d_k|n-m|/\lambda\right)$, where $J_0(\cdot)$ is zero-order Bessel function of the first kind and $d_k$ signifies the inter-antenna distance for the $k$-th user \cite{10163977}. We can obtain $\textbf{G}_{mk}^{\text{d}}\sim\mathcal{CN}(\textbf{0},\mathbf{\Delta}_{mk}^{\text{d}})$ with the assistance of the vectorized channel $\textbf{g}_{mk}^{\text{d}}\in\mathbb{C}^{N_{ap}N_u\times 1}$
\begin{equation}
     \displaystyle \textbf{g}_{mk}^{\text{d}}=\text{vec}(\textbf{G}_{mk}^{\text{d}})=\sqrt{\beta_{mk}}({\textbf{R}}_{k,t}^{1/2} \otimes{\textbf{R}}_{m,r}^{1/2} )\text{vec}(\textbf{v}_{mk}),
     \label{Vector_channel_direct}
   \end{equation}
    where $\otimes$ represents the Kronecker product. Then, the joint correlation matrix
$\mathbf{\Delta}_{mk}^{\text{d}}\in\mathbb{C}^{N_{ap}N_u\times N_{ap}N_u}$ is written as \cite{10201892,10163977}
\begin{equation}
     \displaystyle \mathbf{\Delta}_{mk}^{\text{d}}=\mathbb{E}\{\textbf{g}_{mk}^{\text{d}}(\textbf{g}_{mk}^{\text{d}})^H\}={\beta_{mk}}{\textbf{R}}_{k,t} \otimes{\textbf{R}}_{m,r}.
     \label{Correlation_direct}
   \end{equation}
   
Meanwhile, $\textbf{G}_{mk}^{\text{c}}\in \mathbb{C}^{N_{ap}\times N_{u}}$ represents the cascaded channel from the $k$-th user to the $m$-th AP via the STAR-RIS, which can be expressed as
\begin{equation}
     \displaystyle
     \textbf{G}_{mk}^{\text{c}}=\textbf{G}_{m}\boldsymbol{\Theta}_{\omega_k}\textbf{G}_{k},
    \label{cascaded_uplink_channel_via_RIS}
   \end{equation}
   where 
$\textbf{G}_{m}\in \mathbb{C}^{N_{ap}\times L}$ denotes the channel from the STAR-RIS to the $m$-th AP, given by
\begin{equation}
     \displaystyle \textbf{G}_{m}=\sqrt{\beta_{m}}{\textbf{R}}_{m,r}^{1/2}\textbf{v}_{m}{\textbf{R}}_{m,t}^{{1/2}},
     \label{channel_RIS_AP}
   \end{equation}
where $\beta_{m}$ is the large-scale fading coefficient between the $m$-th AP and STAR-RIS, $\textbf{v}_{m}\in \mathbb{C}^{N_{ap}\times L}$ is composed of i.i.d. $\mathcal{CN}(0,1)$ random variables\cite{10163977}. ${\textbf{R}}_{m,r}  \in \mathbb{C}^{N_{ap}\times N_{ap}}$ denotes the spatial correlation matrix at the $m$-th AP. The spatial correlation matrix for the STAR-RIS is given by ${\textbf{R}}_{m,t} =A{\textbf{R}} \in \mathbb{C}^{L\times L}$, where $A=d_Hd_V$ represents each STAR-RIS element area, with $d_H$ and $d_V$ representing the respective horizontal width and vertical height of each RIS element. The $(x,y)$-th element in $\textbf{R}$ can be obtained by \cite{9598875,9300189}
 \begin{equation}
     \begin{array}{c@{\quad}c}
\displaystyle [\textbf{R}]_{x,y}=\text{sinc}\Bigg{(}\frac{2||\textbf{u}_x-\textbf{u}_y||}{\lambda}\Bigg{)},
     \end{array}
     \label{RIS_spatial_correlation}
   \end{equation}
where $\text{sinc}(a)=\text{sin}(\pi a)/(\pi a)$. $\lambda$ is the carrier wavelength. The position vector is expressed as $\textbf{u}_x=[0,\text{mod}(x-1,L_h)d_h,\lfloor(x-1)/L_h\rfloor d_v]^T$\cite{9598875,10167480,9300189}, where $L_h$ and $L_v$ are the numbers of column elements and row elements at the STAR-RIS, respectively, and $L=L_h\times L_v$.

Moreover, $\boldsymbol{\Theta}_{\omega_k}$ is the STAR-RIS coefficient matrix introduced in Sec. II. A. Additionally, the channel from the $k$-th user to the STAR-RIS is defined as $\textbf{G}_{k}\in \mathbb{C}^{L\times N_u}$, given by
      \begin{equation}
     \begin{array}{c@{\quad}c}
\textbf{G}_{k}=\sqrt{\beta_{k}}{\textbf{R}}_{k,r}^{1/2}\textbf{v}_{k}{\textbf{R}}_{k,t}^{1/2},
     \end{array}
     \label{channel_user_RIS}
   \end{equation}
similar to \eqref{channel_RIS_AP}, $\beta_{k}$ is the large-scale fading coefficient between the $k$-th user and the STAR-RIS. The spatial correlation matrix at the STAR-RIS is denoted as ${\textbf{R}}_{k,r} =A{\textbf{R}} \in \mathbb{C}^{L\times L}$. ${\textbf{R}}_{k,t} \in \mathbb{C}^{N_u\times N_u}$ is the spatial correlation matrix associated with the $k$-th user. $\textbf{v}_{k}\in \mathbb{C}^{L\times N_u}$ is composed of i.i.d. $\mathcal{CN}(0,1)$ random variables \cite{qian2024performance}. 
The cascaded channel can be modelled as $\textbf{G}_{mk}^{\text{c}}\sim\mathcal{CN}(\textbf{0},\mathbf{\Delta}_{mk}^{\text{c}})$. With the assistance of the vectorized channel $\textbf{g}_{mk}^{\text{c}}\in\mathbb{C}^{N_{ap}N_u\times 1}$ presented in \eqref{channel_user_RIS_vec} at the top of this page,
 \begin{figure*}
  \begin{equation}
     \begin{array}{ll}
\displaystyle \textbf{g}_{mk}^{\text{c}}=\text{vec}(\textbf{G}_{mk}^{\text{c}})=\sqrt{\beta_{m}\beta_{k}}\left({\textbf{R}}_{k,t}^{1/2}\otimes {\textbf{R}}_{m,r}^{1/2}\right)\left(\textbf{I}_{N_u}\otimes\textbf{v}_{m}\right)\left(\textbf{I}_{N_u}\otimes(A{\textbf{R}}^{1/2}\boldsymbol{\Theta}_{\omega_k}{\textbf{R}}^{1/2})\right)\text{vec}(\textbf{v}_{k}).
     \end{array}
     \label{channel_user_RIS_vec}
   \end{equation}
   \vspace{-4pt}
   \hrulefill
   \end{figure*}
 we can describe $\mathbf{\Delta}_{mk}^{\text{c}}\in\mathbb{C}^{N_{ap}N_u\times N_{ap}N_u}$ as
 \begin{equation}
     \begin{array}{ll}
\mathbf{\Delta}_{mk}^{\text{c}}=\mathbb{E}\{\textbf{g}_{mk}^{\text{c}}(\textbf{g}_{mk}^{\text{c}})^H\}=\beta_{m}\beta_{k}\text{tr}(\textbf{T}_{\omega_k}){\textbf{R}}_{k,t} \otimes{\textbf{R}}_{m,r},
     \end{array}
     \label{channel_RIS}
   \end{equation}
 with
 \begin{equation}
     \begin{array}{ll}
{\textbf{T}}_{\omega_k}\displaystyle=A^2{\textbf{R}}^{1/2}\boldsymbol{\Theta}_{\omega_k}\textbf{R}\boldsymbol{\Theta}_{\omega_k}^H{\textbf{R}}^{1/2}.
     \end{array}
     \label{T_matrix}
   \end{equation}
Based on \eqref{Correlation_direct}, \eqref{channel_RIS}-\eqref{T_matrix}, the full correlation matrix, $\mathbf{\Delta}_{mk}\in\mathbb{C}^{N_{ap}N_u\times N_{ap}N_u}$ of $\textbf{G}_{mk}\sim\mathcal{CN}(\textbf{0},\mathbf{\Delta}_{mk})$, is
 \begin{equation}
     \begin{array}{ll}
\displaystyle \mathbf{\Delta}_{mk} & \displaystyle=\mathbb{E}\{\textbf{g}_{mk}(\textbf{g}_{mk})^H\}=\mathbf{\Delta}_{mk}^{\text{d}}+\mathbf{\Delta}_{mk}^{\text{c}}\\ & \displaystyle=\left(\beta_{mk}+\beta_{m}\beta_{k}\text{tr}(\textbf{T}_{\omega_k})\right)\left({\textbf{R}}_{k,t} \otimes{\textbf{R}}_{m,r}\right)=\bar{{\Delta}}_{mk}\left({\textbf{R}}_{k,t} \otimes{\textbf{R}}_{m,r}\right),
     \end{array}
\label{channel_correlation}
   \end{equation}
where $\textbf{g}_{mk}=\text{vec}(\textbf{G}_{mk})$,  $\bar{{\Delta}}_{mk}=\beta_{mk}+\beta_{m}\beta_{k}\text{tr}(\textbf{T}_{\omega_k})$ and $\mathbf{\Delta}_{mk}$ are based on the known large-scale fading coefficients and spatial correlation matrices, as previously introduced.

\section{Uplink Channel Estimation with Hardware Impairments}

In this section, we utilize a channel estimation scheme using an uplink channel estimation phase with pilot sequences to investigate channel estimation with hardware impairments \cite{10297571,10163977,10201892}. In the context of uplink channel estimation, $\tau_p$-length pilot sequences are adopted \cite{10201892,10163977}. Notably, $\tau_p$ mutually orthogonal pilot matrices are employed to estimate channels in the multi-antenna ystem. Each pilot matrix is formed by $N_u$ orthogonal $\tau_p$-length pilot sequences drawn from the pilot book \cite{10201892,10163977,9737367}. Thus, the pilot matrix for the $k$-th user is denoted as $\mathbf{\Phi}_{k}\in\mathbb{C}^{\tau_p\times N_u}$. Assigning all users mutually orthogonal pilot matrices is feasible if $\tau_p \geq KN_u$ \cite{8696221,9079911,10571171}. However, in practice, multiple users might share the same pilot matrix with $\tau_p < KN_u$, leading to pilot contamination \cite{qian2024performance,10201892}.
Note that $\lfloor\tau_p/N_u \rfloor$ pilot matrices can be formed to all users \cite{10201892,9737367}, $\tau_p$ should be an integer multiple of $N_u$ to avoid wasting $\tau_p-\lfloor\tau_p/N_u\rfloor$
orthogonal pilot sequences \cite{10163977}. We define $\mathcal{P}_k$ to
denote the user set that shares the same pilot matrix, including the $k$-th user \cite{9737367,10201892,9079911}. Therefore, we can have $\mathbf{\Phi}_{k}^H\mathbf{\Phi}_{k'}= \textbf{I}_{N_u}$ when $ k'\in\mathcal{P}_k$; otherwise, $\mathbf{\Phi}_{k}^H\mathbf{\Phi}_{k'}= \mathbf{0}_{N_u} $ when $ k' \notin \mathcal{P}_k$.
Since we consider hardware impairments in the proposed system, the received signal, $\textbf{Y}_{m,p}\in\mathbb{C}^{N_{ap}\times \tau_p}$, of the $m$-th AP can be defined as
\begin{equation}
\begin{array}{ll}
     \displaystyle \textbf{Y}_{m,p}      \displaystyle =\sqrt{\kappa_{m,r}}\sum\limits_{k=1}^{K}\textbf{G}_{mk}\Big{(}\sqrt{\tau_pp_{p}\kappa_{k,t}}\textbf{P}_{k}^{1/2}\mathbf{\Phi}_{k}^{H}+\textbf{W}_{k,t}^{H}\Big{)}\displaystyle+\textbf{W}_{m,r}+\textbf{N}_{m,p},
\end{array}
\label{Uplink_Received_Pilot}
   \end{equation}
where $p_p$ corresponds to the pilot transmit power of each user. $\textbf{P}_{k}=\text{diag}(\xi_{k1},...,\xi_{kN_u})$ represents the power control matrix of the $k$-th user, $0\leq\xi_{kn}\leq 1$ is the power control factor of the $n$-th antenna on the $k$-th user \cite{10163977,10201892,9737367}. Moreover, $\textbf{N}_{m,p}\in\mathbb{C}^{N_{ap}\times \tau_p}$ is the additive white Gaussian noise (AWGN) matrix with the $v$-th column satisfying $\Big{[}\textbf{N}_{m,p}\Big{]}_v\sim \mathcal{CN}(\textbf{0},\sigma^2\textbf{I}_{N_{ap}})$. 
{\color{black}To introduce the non-ideal hardware at the transceiver, we model the hardware impairments as
mutually uncorrelated Gaussian random variables
since the aggregate contribution of many impairments results in these transmitter and receiver distortions \cite{9528977,8891922}.}
In this case, $\textbf{W}_{k,t}\in\mathbb{C}^{\tau_p\times N_u}$ refers to the transmitter distortion at the $k$-th user and $\textbf{W}_{m,r}\in\mathbb{C}^{N_{ap}\times \tau_p}$ denotes the receiver distortion at the $m$-th AP. According to \cite{10225319,9459571,10163977}, we can have 
\begin{equation}
\begin{array}{ll}
     \displaystyle \textbf{W}_{k,t}   \sim  \mathcal{CN}\Big{(}\textbf{0},~(1-\kappa_{k,t})p_p\textbf{P}_{k}\otimes \textbf{I}_{\tau_p}\Big{)},
\end{array}
\label{HWI_transmitter}
   \end{equation}
     \begin{equation}
\begin{array}{ll}
     \displaystyle \textbf{W}_{m,r} \sim  \mathcal{CN}\Big{(}\textbf{0},\textbf{I}_{\tau_p}\otimes \textbf{C}_{m|\{\textbf{G}_{mk}\}}\Big{)}, 
\end{array}
\label{HWI_AP}
   \end{equation}
where \eqref{HWI_AP} presents the conditional distribution based on the set of channel vectors $\{\textbf{G}_{mk}\}$ within a given coherence interval. Additionally, the hardware impairment parameters, $\kappa_{m,r}\in[0,1],~\forall m$ and $\kappa_{k,t}\in[0,1],~\forall k$ describe the respective hardware quality of the transmitter and receiver\cite{10225319,10201892}. Note that $\kappa_{m,r}=\kappa_{k,t}=1$ represents the perfect transceiver and $\kappa_{m,r}=\kappa_{k,t}=0$ is when there is no transceiver \cite{10225319}.
The conditional covariance can be expressed as \cite{10163977,10201892}
     \begin{equation}
\begin{array}{ll}
     \displaystyle\textbf{C}_{m|\{\textbf{G}_{mk}\}}&\displaystyle=(1-\kappa_{m,r})p_p\sum\limits_{k=1}^K\text{diag}\left( |[\textbf{G}_{mk}]_{1}\textbf{P}_k^{\frac{1}{2}}|^2,...,|\textbf{G}_{mk}]_{N_{ap}}\textbf{P}_k^{\frac{1}{2}}|^2\right)\\&\displaystyle=(1-\kappa_{m,r})p_p\sum\limits_{k=1}^K\textbf{G}_{mk}\textbf{P}_k\textbf{G}_{mk}^H\odot\textbf{I}_{N_{ap}}\\&\displaystyle=(1-\kappa_{m,r})p_p\sum\limits_{k=1}^K
 \bar{{\Delta}}_{mk}\text{tr}\left(\textbf{P}_k{\textbf{R}}_{k,t}\right){\textbf{R}}_{m,r}\odot\textbf{I}_{N_{ap}},
\end{array}
\label{HWI_receiver}
   \end{equation}
where $[\textbf{G}_{mk}]_{n}$ is the $n$-th row of channel matrix $\textbf{G}_{mk}$, $\odot$ represents the element-wise product.

Then, the $m$-th AP multiplies $\textbf{Y}_{m}$ with $\mathbf{\Phi}_{k}$ to obtain the projection as \eqref{Pilot_projection} at the top of this page \cite{qian2024performance}.
\begin{figure*}
     \begin{equation}
\begin{array}{ll}
     \displaystyle \textbf{Y}_{mk,p}\displaystyle=\textbf{Y}_{m,p}\mathbf{\Phi}_{k}&\displaystyle=\sqrt{\kappa_{m,r}}\sum\limits_{k'=1}^{K}\textbf{G}_{mk'}\Big{(}\sqrt{\tau_pp_{p}\kappa_{k',t}}\textbf{P}_{k'}^{1/2}\mathbf{\Phi}_{k'}^{H}+\textbf{W}_{k',t}^{H}\Big{)}\mathbf{\Phi}_{k}+\textbf{W}_{m,r}\mathbf{\Phi}_{k}+\textbf{N}_{m,p}\mathbf{\Phi}_{k}\\ &\displaystyle=\sqrt{\kappa_{m,r}}\sum\limits_{k'\in\mathcal{P}_k}\sqrt{\tau_pp_{p}\kappa_{k',t}}\textbf{G}_{mk'}\textbf{P}_{k'}^{1/2}+\sqrt{\kappa_{m,r}}\sum\limits_{k'=1}^{K}\textbf{G}_{mk'}\textbf{W}_{k',t}^{H}\mathbf{\Phi}_{k}+\textbf{W}_{m,r}\mathbf{\Phi}_{k}+\textbf{N}_{m,p}\mathbf{\Phi}_{k}.
\end{array}
\label{Pilot_projection}
   \end{equation}
   \vspace{-4 pt}
   \hrulefill
   \end{figure*} 
Based on the vectorization
operation for $\textbf{Y}_{mk,p}$, we can have $\textbf{y}_{mk,p}=\text{vec}(\textbf{Y}_{mk,p})$ following \eqref{Pilot_projection_vectorized} at the top of this page.
\begin{figure*}
\vspace{-4 pt}
 \begin{equation}
\begin{array}{ll}
     \displaystyle \textbf{y}_{mk,p}&\displaystyle=\text{vec}(\textbf{Y}_{mk,p})\\ &\displaystyle=\sqrt{\kappa_{m,r}}\sum\limits_{k'\in\mathcal{P}_k}\sqrt{\tau_pp_{p}\kappa_{k',t}}\text{vec}(\textbf{G}_{mk'}\textbf{P}_{k'}^{1/2})+\sqrt{\kappa_{m,r}}\sum\limits_{k'=1}^{K}\text{vec}(\textbf{G}_{mk'}\textbf{W}_{k',t}^{H}\mathbf{\Phi}_{k})+\text{vec}(\textbf{W}_{m,r}\mathbf{\Phi}_{k})+\text{vec}(\textbf{N}_{m,p}\mathbf{\Phi}_{k})\\ &\displaystyle=\sqrt{\kappa_{m,r}}\sum\limits_{k'\in\mathcal{P}_k}\sqrt{\tau_pp_{p}\kappa_{k',t}}\left(\textbf{P}_{k'}^{1/2}\otimes\textbf{I}_{N_{ap}}\right)\textbf{g}_{mk'}+\sqrt{\kappa_{m,r}}\sum\limits_{k'=1}^{K}\left((\textbf{W}_{k',t}^{H}\mathbf{\Phi}_{k})^T\otimes\textbf{I}_{N_{ap}}\right)\textbf{g}_{mk'}\displaystyle+\left(\mathbf{\Phi}_{k}^T\otimes\textbf{I}_{N_{ap}}\right)\left(\text{vec}(\textbf{W}_{m,r})+\text{vec}(\textbf{N}_{m,p})\right),
\end{array}
\label{Pilot_projection_vectorized}
   \end{equation}
   \vspace{-4 pt}
\hrulefill
   \end{figure*}
Then, we can compute the MMSE channel estimation of ${\textbf{G}}_{mk}$ using vectorization $\hat{\textbf{g}}_{mk}=\text{vec}(\hat{\textbf{G}}_{mk})$ \cite{10163977,9079911,10001172}, given by
  \begin{equation}
\begin{array}{ll}
     \displaystyle \hat{\textbf{g}}_{mk}=\text{vec}(\hat{\textbf{G}}_{mk})=\mathbf{Q}_{mk}\mathbf{\Psi}_{mk}^{-1}\textbf{y}_{mk,p},
\end{array}
   \end{equation}
where
 \begin{equation}
\begin{array}{ll}
     \displaystyle \mathbf{Q}_{mk}\displaystyle=\mathbb{E}\left\{ {\textbf{g}}_{mk}\textbf{y}_{mk,p}^H\right\}=\sqrt{\tau_pp_{p}\kappa_{m,r}\kappa_{k,t}}\bar{\mathbf{\Delta}}_{mk}\Big{(}\textbf{R}_{k,t}\textbf{P}_{k}^{1/2}\otimes\textbf{R}_{m,r}\Big{)},
\end{array}
\label{Q_mk}
   \end{equation}
   $\mathbf{\Psi}_{mk}$ is expressed as \eqref{Psi_mk} at the top of this page.
   \begin{figure*}
   \vspace{-4 pt}
 \begin{equation}
\begin{array}{ll}
     \displaystyle \mathbf{\Psi}_{mk}\displaystyle=\mathbb{E}\left\{ {\textbf{y}}_{mk}\textbf{y}_{mk,p}^H\right\}&\displaystyle=\kappa_{m,r}\sum\limits_{k'\in\mathcal{P}_k}\tau_pp_{p}\kappa_{k',t}\bar{\mathbf{\Delta}}_{mk'}\bigg{(}\Big{(}\textbf{P}_{k'}^{1/2}\textbf{R}_{k',t}\textbf{P}_{k'}^{1/2}\Big{)}\otimes\textbf{R}_{m,r}\bigg{)}\\&\displaystyle+\kappa_{m,r}\sum\limits_{k'=1}^K(1-\kappa_{k',t})p_p\bar{\mathbf{\Delta}}_{mk'}\text{tr}\Big{(}\textbf{P}_{k'}\textbf{R}_{k',t}\Big{)}\left(\textbf{I}_{N_u}\otimes\textbf{R}_{m,r}\right)+
     \left(\textbf{I}_{N_u}\otimes\textbf{C}_{m|\{\textbf{G}_{mk}\}}\right)+ \sigma^2\left(\textbf{I}_{N_u}\otimes\textbf{I}_{N_{ap}}\right) .
\end{array}
   \label{Psi_mk}
   \end{equation}
    \vspace{-4 pt}
\hrulefill
   \end{figure*}
Then, the estimate $\hat{\textbf{G}}_{mk}
$ and the estimation error $\tilde{\textbf{G}}_{mk}
={\textbf{G}}_{mk}
-\hat{\textbf{G}}_{mk}
$ with $\tilde{\textbf{g}}_{mk}=\text{vec}(\tilde{\textbf{G}}_{mk})$, are distributed as $\mathcal{CN}\left(\textbf{0},\hat{\mathbf{\Delta}}_{mk}\right)$ and $\mathcal{CN}\left(\textbf{0},\mathbf{\Delta}_{mk}-\hat{\mathbf{\Delta}}_{mk}\right)$, respectively, where
   \begin{equation}
\begin{array}{ll}
     \displaystyle \hat{\mathbf{\Delta}}_{mk}=\textbf{Q}_{mk}(\textbf{Q}_{mk}\mathbf{\Psi}_{mk}^{-1})^H.
     \end{array}
     \label{Num_direct}
   \end{equation}
   Note that $\tilde{\textbf{G}}_{mk}$ is uncorrelated with $\hat{\textbf{G}}_{mk}$ but remains dependent on it due to the non-Gaussian characteristics of the channel estimates\cite{10163977}.
  \\
\textit{Proof}: Please refer to Appendix \eqref{Appendix_NMSE}.
    
\section{Uplink Data Transmission and Performance Analysis}

This section investigates the achievable
uplink SE in hardware-impaired scenarios utilizing the MMSE-SIC detector \cite{10163977,9737367}. Attributable to the adaptable network topology of cell-free massive MIMO, four signal processing implementations were studied in \cite{8845768,9737367}. As far as we know, most existing analyses of RIS-assisted cell-free massive MIMO systems have considered the classical processing implementation (Local Processing and Centralized Decoding) \cite{10225319,9322151,10167480}, similar to STAR-RIS-assisted cell-free massive MIMO systems \cite{qian2024performance,10264149}. Therefore, there is a need to investigate the advanced level of cell-free massive MIMO operation (fully centralized processing) in STAR-RIS-assisted cell-free massive MIMO systems with multi-antenna users to provide practical guidelines and insights. In this case, we focus on two main uplink implementations: local processing and centralized decoding (Level 1) and fully centralized processing (Level 2). The comparison of these two levels is explored below.

\subsection{Level 1: Local Processing and Centralized
Decoding}
This implementation is treated as a two-layer decoding scheme. Initially, all user antennas simultaneously transmit their data symbols to the APs \cite{8696221,9737367}. For the $k$-th user, the transmitted data streams are denoted by $\textbf{s}_k=[s_{k1},...,s_{kN_u}]^T$, where $s_{kn}\sim\mathcal{CN}(0,1),~\forall n$. The power control matrix of the $k$-th user is $\textbf{P}_k$. Subsequently, the received signal at the $m$-th AP, $\textbf{y}_m\in\mathbb{C}^{N_{ap}\times 1}$, experiencing the hardware impairments, can be given by
\begin{equation}
\begin{array}{ll}
     \displaystyle \textbf{y}_m=\sqrt{\kappa_{m,r}}\sum\limits_{k=1}^{K}\textbf{G}_{mk}\Big{(}\sqrt{p_u\kappa_{k,t}}\textbf{P}_{k}^{1/2}\textbf{s}_{k}+\boldsymbol{\eta}_{k,t}^H\Big{)}\displaystyle+\boldsymbol{\eta}_{m,r}+\textbf{n}_{m},
\end{array}\label{y_m}
   \end{equation}
where $\boldsymbol{\eta}_{k,t}\in\mathbb{C}^{1\times N_u}$ denotes the transmission distortion of the $k$-th user following $\mathcal{CN}\Big{(}\textbf{0},~(1-\kappa_{k,t})p_u\textbf{P}_{k}\Big{)}$. $\boldsymbol{\eta}_{m,r}\in\mathbb{C}^{N_{ap}\times 1}$ is the receiver distortion of the $m$-th AP with $\mathcal{CN}\Big{(}\textbf{0},\bar{\textbf{C}}_{m|\{\textbf{G}_{mk}\}}\Big{)}$. Similar to \eqref{HWI_receiver}, $\bar{\textbf{C}}_{m|\{\textbf{G}_{mk}\}}=(1-\kappa_{m,r})p_u\sum\nolimits_{k=1}^K     \left(\beta_{mk}+\beta_{m}\beta_{k}\text{tr}(\textbf{T}_{\omega_k})\right)\text{tr}\left(\textbf{P}_k{\textbf{R}}_{k,t}\right){\textbf{R}}_{m,r}\odot\textbf{I}_{N_{ap}}$ \cite{8476516,8171057}. $\textbf{n}_{m}\sim\mathcal{CN}(0,\sigma^2\textbf{I}_{N_{ap}})$ is the AWGN at the $m$-th AP.
   
Subsequently, the received signal $\textbf{y}_{m}$ at the $m$-th AP is multiplied with the combining
matrix $\textbf{V}_{mk}\in\mathbb{C}^{N_{ap}\times N_u}$ to detect the symbol from the $k$-th user\cite{9665300}. Sequentially, the obtained quantity $\check{\textbf{s}}_{mk}\triangleq {\textbf{V}}_{mk}^H\textbf{y}_m$ is sent to the CPU. Then, the CPU uses the second layer decoding, the weight matrix $\textbf{A}_{mk}\in\mathbb{C}^{N_u\times N_u}$, with $\textbf{A}_{k}=[\textbf{A}_{1k}^T,\textbf{A}_{2k}^T,...,\textbf{A}_{Mk}^T]^T\in\mathbb{C}^{MN_u\times N_u}$, to obtain $\hat{\textbf{s}}_{k}\in\mathbb{C}^{N_u\times 1}$ \cite{9737367} as \eqref{CPU_received_signal} at the top of this page. In \eqref{CPU_received_signal}, $\text{DS}_{{k}}$ is the desired signal, $\text{BU}_{{k}}$ is the beamforming gain uncertainty, $\text{IU}_{{kk}'}$ is the inter-user interference, $\text{TD}_{{kk'}}$ is the transmitter distortion, $\text{RD}_{{k}}$ is the receiver distortion and $\text{NS}_{{k}}$ is the noise term, respectively.
\begin{figure*}[!t]
 \begin{equation}
\begin{array}{ll}
     \displaystyle \hat{\textbf{s}}_{k}& \displaystyle=\sum\limits_{m=1}^M \textbf{A}^H_{mk}\check{\textbf{s}}_{mk}=\sum\limits_{m=1}^M \textbf{A}^H_{mk}{\textbf{V}}_{mk}^H\textbf{y}_m\\& \displaystyle=\underbrace {\sum\limits_{m=1}^M\sqrt{\kappa_{m,r}p_u\kappa_{k,t}}\textbf{A}^H_{mk}\mathbb{E}\left\{{\textbf{V}}_{mk}^H{\textbf{G}}_{mk}{\textbf{P}}_{k}^{1/2}\right\}\textbf{s}_k}_{\text{DS}_k}+\underbrace{\sum\limits_{m=1}^M\sqrt{\kappa_{m,r}p_u\kappa_{k,t}}\textbf{A}^H_{mk}\bigg{(}{\textbf{V}}_{mk}^H{\textbf{G}}_{mk}{\textbf{P}}_{k}^{1/2}-\mathbb{E}\left\{{\textbf{V}}_{mk}^H{\textbf{G}}_{mk}{\textbf{P}}_{k}^{1/2}\right\}\bigg{)}\textbf{s}_k}_{\text{BU}_k}\\&\displaystyle+\sum\limits_{k'\neq k}^K \underbrace{\sum\limits_{m=1}^M\sqrt{\kappa_{m,r}p_u\kappa_{k',t}}\textbf{A}^H_{mk}{\textbf{V}}_{mk}^H{\textbf{G}}_{mk'}{\textbf{P}}_{k'}^{1/2}\textbf{s}_{k'}}_{\text{IU}_{kk'}}+\sum\limits_{k'=1}^K\underbrace{\sum\limits_{m=1}^M\sqrt{\kappa_{m,r}}\textbf{A}^H_{mk}{\textbf{V}}_{mk}^H{\textbf{G}}_{mk'}\boldsymbol{\eta}_{k',t}^H}_{\text{TD}_{kk'}}+\underbrace{\sum\limits_{m=1}^M \textbf{A}^H_{mk}{\textbf{V}}_{mk}^H\boldsymbol{\eta}_{m,r}}_{\text{RD}_k}+\underbrace{\sum\limits_{m=1}^M \textbf{A}^H_{mk}{\textbf{V}}_{mk}^H\textbf{n}_{m}}_{\text{NS}_k}.
\end{array}
   \label{CPU_received_signal}
   \end{equation}
    \vspace{-8pt}
   \hrulefill
   \end{figure*}
Consequently, based on \eqref{CPU_received_signal}, the lower bound of the uplink achievable SE of the $k$-th user with the
MMSE-SIC detector \cite{10163977,10201892} can be determined by
\begin{equation}
\begin{array}{ll}
     \displaystyle \text{SE}_k=\frac{\tau_c-\tau_p}{\tau_c}\text{log}_2\Big{|}\textbf{I}_{N_u}+\textbf{D}_{{k}}^H\boldsymbol{\Sigma}_k^{-1}\textbf{D}_{{k}}\Big{|}.
\end{array}\label{uplink_SE_level1}
   \end{equation}
Note that a pre-log factor arises because only $\tau_c-\tau_p$ symbols are used for data transmission\cite{10201892}. Moreover, $\textbf{D}_{{k}}$ is given by
   \begin{equation}
\begin{array}{ll}
     \displaystyle {\textbf{D}}_{k}=\sqrt{p_u}{\textbf{A}}_{k}^H\mathbb{E}\Big{\{}\textbf{H}_{kk}\Big{\}}\textbf{P}_{k}^{1/2},
\end{array}\label{D_k}
   \end{equation}
   and $\boldsymbol{\Sigma}_k$ is determined by \eqref{Sigma_k} at the top of this page.
   \begin{figure*}[t!]
   \vspace{-4 pt}
 \begin{equation}
\begin{array}{ll}
     \displaystyle \boldsymbol{\Sigma}_k&\displaystyle=\underbrace{\sum\limits_{k'=1}^Kp_u{\textbf{A}}_{k}^H\mathbb{E}\Big{\{}\textbf{H}_{kk'}\textbf{P}_{k'}\textbf{H}_{kk'}^H\Big{\}}\textbf{A}_{k}-{\textbf{D}}_{k}{\textbf{D}}_{k}^H}_{\text{IU}_k+\text{BU}_k}\displaystyle+\underbrace{\sum\limits_{k'=1}^Kp_u\Bigg{(}{\textbf{A}}_{k}^H\mathbb{E}\Big{\{}\textbf{F}_{kk'}\textbf{P}_{k'}\textbf{F}_{kk'}^H\Big{\}}\textbf{A}_{k}-{\textbf{A}}_{k}^H\mathbb{E}\Big{\{}\textbf{H}_{kk'}\textbf{P}_{k'}\textbf{H}_{kk'}^H\Big{\}}\textbf{A}_{k}\Bigg{)}}_{\text{TD}_k}\displaystyle+\underbrace{\textbf{A}_{k}^H\mathbf{\Gamma}_k\textbf{A}_{k}}_{\text{RD}_k}+\underbrace{\sigma^2\textbf{A}_{k}^H\mathbf{\Lambda}_k\textbf{A}_{k}}_{\text{NS}_k}\\ &\displaystyle=\sum\limits_{k'=1}^Kp_u{\textbf{A}}_{k}^H\mathbb{E}\Big{\{}\textbf{F}_{kk'}\textbf{P}_{k'}\textbf{F}_{kk'}^H\Big{\}}\textbf{A}_{k}-{\textbf{D}}_{k}{\textbf{D}}_{k}^H+\textbf{A}_{k}^H\mathbf{\Gamma}_k\textbf{A}_{k}+\sigma^2\textbf{A}_{k}^H\mathbf{\Lambda}_k\textbf{A}_{k}.
\end{array}
\label{Sigma_k}
\vspace{-4 pt}
   \end{equation}
    \vspace{-8pt}
   \hrulefill
   \end{figure*}
We simplify the expressions by defining ${\textbf{H}}_{kk'}\in\mathbb{C}^{MN_u\times N_u}$, ${\textbf{F}}_{kk'}\in\mathbb{C}^{MN_u\times N_u}$, ${\mathbf{\Gamma}}_{k}\in\mathbb{C}^{MN_u\times MN_u}$ and ${\mathbf{\Lambda}}_{k}\in\mathbb{C}^{MN_u\times MN_u}$  as
 \begin{equation}
\begin{array}{ll}
     \displaystyle {\textbf{H}}_{kk'}=[(\sqrt{\kappa_{1,r}\kappa_{k',t}}{\textbf{V}}_{1k}^H{\textbf{G}}_{1k'})^T,...,(\sqrt{\kappa_{M,r}\kappa_{k',t}}{\textbf{V}}_{Mk}^H{\textbf{G}}_{Mk'})^T]^T,
\end{array}\label{H_kk}
   \end{equation}
   \begin{equation}
\begin{array}{ll}
     \displaystyle {\textbf{F}}_{kk'}=[(\sqrt{\kappa_{1,r}}{\textbf{V}}_{1k}^H{\textbf{G}}_{1k'})^T,...,(\sqrt{\kappa_{M,r}}{\textbf{V}}_{Mk}^H{\textbf{G}}_{Mk'})^T]^T,
\end{array}\label{F_kk}
   \end{equation}
     \begin{equation}
\begin{array}{ll}
     \displaystyle {\mathbf{\Gamma}}_{k}=\text{blkdiag}\Big{(}
     \mathbb{E}\left\{\textbf{V}_{1k}^H\bar{\textbf{C}}_{1|\{\textbf{G}_{mk}\}}\textbf{V}_{1k}\right\},...,\mathbb{E}\left\{\textbf{V}_{Mk}^H\bar{\textbf{C}}_{M|\{\textbf{G}_{mk}\}}\textbf{V}_{Mk}\right\}
     \Big{)},
\end{array}\label{Gamma_kk}
   \end{equation}
        \begin{equation}
\begin{array}{ll}
     \displaystyle {\mathbf{\Lambda}}_{k}=\text{blkdiag}\Big{(}
     \mathbb{E}\left\{\textbf{V}_{1k}^H\textbf{V}_{1k}\right\},...,\mathbb{E}\left\{\textbf{V}_{Mk}^H\textbf{V}_{Mk}\right\}
     \Big{)}.
\end{array}\label{Lambda_kk}
   \end{equation}

Note that the above formulas hold for any combining vector, with the $m$-th AP, leveraging its local estimate $\hat{\textbf{G}}_{mk}$ to design ${\textbf{V}}_{mk}$. A potential choice is MR combining $\textbf{V}_{mk}=\hat{\textbf{G}}_{mk}$. Besides, the local MMSE combining matrix, designed to minimize the mean-square error (MSE) $\text{MSE}_{mk}=\mathbb{E}\{||\textbf{s}_k- {\textbf{V}}_{mk}^H\textbf{y}_m||^2|\hat{\textbf{G}}_{mk}\}$ is introduced as \eqref{Local_MMSE} at the top of this page
\begin{figure*}
      \begin{equation}
\begin{array}{ll}
     \displaystyle {\textbf{V}}_{mk}=\left[
     \begin{array}{ll}
     \sum\limits_{k'=1}^K \kappa_{m,r}p_u\left(\hat{\textbf{G}}_{mk'}\textbf{P}_{k'}\hat{\textbf{G}}_{mk'}^H+\tilde{\textbf{C}}_{mk'}\right)+\bar{\textbf{C}}_{m|\textbf{G}_{mk}}+\sigma^2\textbf{I}_{N_{ap}}      
\end{array}\right]^{-1}\sqrt{\kappa_{m,r}p_u\kappa_{k,t}}\hat{\textbf{G}}_{mk}\textbf{P}_k^{1/2},
\end{array}\label{Local_MMSE}
   \end{equation}
   \vspace{-8pt}
   \hrulefill
   \end{figure*}
where
 \begin{equation}
\begin{array}{ll}
     \displaystyle \tilde{\textbf{C}}_{mk}=\mathbb{E}\Big{\{}\tilde{\textbf{G}}_{mk}\textbf{P}_k\tilde{\textbf{G}}_{mk}^H\Big{\}}=\sum\limits_{n=1}^{N_u}\xi_{kn}\left({\boldsymbol{\Delta}}_{mk}^{n,n}-\hat{\boldsymbol{\Delta}}_{mk}^{n,n}\right).
\end{array}
   \end{equation}
   \textit{Proof:} The proof follows straightforwardly from the standard matrix derivation results in \cite{10201892,8845768} and is omitted for brevity.

\textit{Remark 1:} The design of local MMSE combining aims to minimize MSE and maximize uplink SE using local CSI at each AP. The Monte Carlo method allows for the computation of achievable SE expressions. However, deriving closed-form expressions for the achievable SE with local MMSE combining is hard due to the analytical intractability of inverting random matrices \cite{9737367,10201892}.

\subsubsection{Large-Scale Fading Decoding (LSFD)} For the second layer decoding, the CPU designs the complex LSFD coefficient matrix ${\textbf{A}}_{k}$ based on channel statistics \cite{qian2024performance}. This matrix maximizes the achievable SE for the $k$-th user \cite{8845768} and can be given by
\begin{equation}
\begin{array}{ll}
     \displaystyle {\textbf{A}}_{k}=\bigg{(}p_u\sum\limits_{k'=1}^K\mathbb{E}\Big{\{}
     \textbf{F}_{kk'}\textbf{P}_{k'}\textbf{F}_{kk'}^H\Big{\}}+\mathbf{\Gamma}_k+\sigma^2\mathbf{\Lambda}_k\bigg{)}^{-1}\mathbb{E}\Big{\{}\textbf{H}_{kk}\Big{\}}\textbf{P}_{k}^{1/2}.
\end{array}\label{LSFD}
   \end{equation}
 
\subsubsection{Simple Centralized Decoding}
According to the above observations, the LSFD method can maximize SE at the cost of high complexity \cite{qian2024performance}. For the sake of ease, the CPU can alternatively take the Matched Filter (MF) method $\textbf{A}_{mk}=\frac{1}{M}\textbf{I}_{N_u}\in\mathbb{C}^{N_u\times N_u}$ to weight the local estimates, this is considered as the so-called simple centralized decoding \cite{8845768,9737367}.

\textit{Theorem 1}: By adopting MR combining $\textbf{V}_{mk}=\hat{\textbf{G}}_{mk}$ at APs \cite{10201892}, we can deliver the closed-form SE expressions of Level 1 with MMSE-SIC detector as
\begin{equation}
\begin{array}{ll}
     \displaystyle \text{SE}_k=\frac{\tau_c-\tau_p}{\tau_c}\text{log}_2\Big{|}\textbf{I}_{N_u}+\bar{\textbf{D}}_{{k}}^H\bar{\boldsymbol{\Sigma}}_k^{-1}\bar{\textbf{D}}_{{k}}\Big{|}.
\end{array}\label{uplink_SE}
   \end{equation}
where 
\begin{equation}
\begin{array}{ll}
     \displaystyle \bar{\textbf{D}}_{k}= \sqrt{p_u}{\textbf{A}}_{k}^H\bar{\textbf{H}}_{kk}\textbf{P}_{k}^{1/2},
\end{array}\label{D_k_CF}
   \end{equation}
   \begin{equation}
\begin{array}{ll}
     \displaystyle \bar{\boldsymbol{\Sigma}}_k&\displaystyle=\sum\limits_{k'=1}^Kp_u{\textbf{A}}_{k}^H\textbf{U}_{kk'}\textbf{A}_{k}-\bar{\textbf{D}}_{k}\bar{\textbf{D}}_{k}^H+\textbf{A}_{k}^H\bar{\mathbf{\Gamma}}_k\textbf{A}_{k}+\sigma^2\textbf{A}_{k}^H\bar{\mathbf{\Lambda}}_k\textbf{A}_{k}.
\end{array}
\label{Sigma_k_CF}
   \end{equation}

\textit{Proof:} Please see Appendix \eqref{Appendix_SE}.
\subsection{Level 2: Fully Centralized Processing}
Fully Centralized Processing can be treated as the most advanced level of cell-free massive MIMO operation \cite{8845768}. The CPU processes the
channel estimation and data signal detection, and the APs serve as relays to forward all signals to the CPU \cite{8845768,9737367,10326460}.
The pilot signal $\textbf{Y}_{p}\in\mathbb{C}^{MN_{ap}\times \tau_p}$ at the CPU is defined as \eqref{Uplink_Received_Pilot_CPU} at the top of this page.
\begin{figure*}[t!]
\begin{equation}
\begin{array}{ll}
     \displaystyle \textbf{Y}_{p}      \displaystyle =\left[\begin{array}{cc}
         \textbf{Y}_{1,p}\\\vdots\\\textbf{Y}_{M,p}
     \end{array}
     \right]=
     \sum\limits_{k=1}^{K}\left[\begin{array}{cc}
        \sqrt{\kappa_{1,r}}\textbf{G}_{1k}\\\vdots\\
        \sqrt{\kappa_{M,r}}\textbf{G}_{Mk}
     \end{array}
     \right]\Big{(}\sqrt{\tau_pp_{p}\kappa_{k,t}}\textbf{P}_{k}^{1/2}\mathbf{\Phi}_{k}^{H}+\textbf{W}_{k,t}^{H}\Big{)}+\left[\begin{array}{cc}
         \textbf{W}_{1,r}\\\vdots\\\textbf{W}_{M,r}
     \end{array}
     \right]+\left[\begin{array}{cc}
         \textbf{N}_{1,p}\\\vdots\\\textbf{N}_{M,p}
     \end{array}
     \right].
\end{array}
\label{Uplink_Received_Pilot_CPU}
   \end{equation}
    \vspace{-8pt}
   \hrulefill
\end{figure*}

The equivalent collective channel $\textbf{g}_k\in\mathbb{C}^{{MN_{ap}N_u}\times 1}$ for the $k$-th user is given by $\textbf{g}_k=\left[
\text{vec}(\sqrt{\kappa_{1,r}}\textbf{G}_{1k})^T,...,\text{vec}(\sqrt{\kappa_{M,r}}\textbf{G}_{Mk})^T
\right]^T\sim\mathcal{CN}\left(\textbf{0},{\mathbf{\Delta}}_{k}\right)$, where ${\mathbf{\Delta}}_{k}=\text{blkdiag}\left(\kappa_{1,r}{\mathbf{\Delta}}_{1k},...,\kappa_{M,r}{\mathbf{\Delta}}_{Mk}
\right)\in\mathbb{C}^{{MN_{ap}N_u}\times {MN_{ap}N_u}}$ is the entire collective correlation matrix of the $k$-th user. The CPU computes all MMSE estimates like the local estimation \eqref{Pilot_projection} and \eqref{Pilot_projection_vectorized}. The collective channel estimate $\hat{\textbf{g}}_{k} \in\mathbb{C}^{{MN_{ap}N_u}\times 1}$ for the $k$-th user with $\hat{\textbf{G}}_{k}=\left[
(\sqrt{\kappa_{1,r}}\hat{\textbf{G}}_{1k})^T,...,(\sqrt{\kappa_{M,r}}\hat{\textbf{G}}_{Mk})^T
\right]^T\in\mathbb{C}^{{MN_{ap}}\times N_u}$, can be formed as 
\begin{equation}
\begin{array}{ll}
     \displaystyle \hat{\textbf{g}}_{k}& \displaystyle=\left[
\text{vec}(\sqrt{\kappa_{1,r}}\hat{\textbf{G}}_{1k})^T,...,\text{vec}(\sqrt{\kappa_{M,r}}\hat{\textbf{G}}_{Mk})^T
\right]^T\\ &\displaystyle=\left[(\sqrt{\kappa_{1,r}}\hat{\textbf{g}}_{1k})^T,...,(\sqrt{\kappa_{M,r}}\hat{\textbf{g}}_{Mk})^T
\right]^T,
\end{array}
   \end{equation}
where $ \hat{\textbf{g}}_{k} \sim\mathcal{CN}\left(\textbf{0},\hat{\mathbf{\Delta}}_{k}\right)$ with $\hat{\mathbf{\Delta}}_{k}=\textbf{Q}_k\mathbf{\Psi}_{k}^{-1}\textbf{Q}_k^H$. $\textbf{Q}_k=\text{blkdiag}\left(\sqrt{\kappa_{1,r}}\textbf{Q}_{1k},...\sqrt{\kappa_{M,r}}\textbf{Q}_{Mk}
\right)$ and $\mathbf{\Psi}_{k}^{-1}=\text{blkdiag}\left(\mathbf{\Psi}_{1k}^{-1},...\mathbf{\Psi}_{Mk}^{-1}
\right)$.

Then, the received signal at the CPU for data detection is expressed as \eqref{y_CPU} at the top of this page 
\begin{figure*}[t!]
\begin{equation}
\begin{array}{ll}
     \displaystyle \underbrace{[\textbf{y}_1^T,...,\textbf{y}_M^T]^T}_{=\textbf{y}\in\mathbb{C}^{{MN_{ap}\times1}}}
=\sum\limits_{k=1}^{K}\underbrace{[(\sqrt{\kappa_{1,r}}\textbf{G}_{1k})^T,...,(\sqrt{\kappa_{M,r}}\textbf{G}_{Mk})^T]^T}_{=\textbf{G}_k\in\mathbb{C}^{{MN_{ap}\times N_u}}}
\Big{(}\sqrt{p_u\kappa_{k,t}}\textbf{P}_{k}^{1/2}\textbf{s}_{k}+\boldsymbol{\eta}_{k,t}^H\Big{)}\displaystyle+\underbrace{[\boldsymbol{\eta}_{1,r}^T,...,\boldsymbol{\eta}_{M,r}^T]^T}_{=\boldsymbol{\eta}_r\in\mathbb{C}^{{MN_{ap}\times1}}}+\underbrace{[\textbf{n}_{1}^T,...,\textbf{n}_{M}^T]^T}_{=\textbf{n}\in\mathbb{C}^{{MN_{ap}\times1}}},
\end{array}\label{y_CPU}
   \end{equation}
   \vspace{-8pt}
   \hrulefill
\end{figure*}
or in the compact form as
\begin{equation}
\begin{array}{ll}
     \displaystyle \textbf{y}
=\sum\limits_{k=1}^{K}\textbf{G}_k
\Big{(}\sqrt{p_u\kappa_{k,t}}\textbf{P}_{k}^{1/2}\textbf{s}_{k}+\boldsymbol{\eta}_{k,t}^H\Big{)}\displaystyle+\boldsymbol{\eta}_r+\textbf{n}.
\end{array}\label{y_CPU_compact}
   \end{equation}
Next, the CPU selects an arbitrary combining matrix
$\textbf{V}_k \in\mathbb{C}^{MN_{ap}\times N_u}$ depending on the collective channel estimates to detect $\textbf{s}_{k}$ as
\begin{equation}
\begin{array}{ll}
     \displaystyle \check{\textbf{s}}_{k}&   \displaystyle =\textbf{V}_k^H\textbf{y}
\\ &\displaystyle=\sum\limits_{k=1}^{K}\textbf{V}_k^H\textbf{G}_k
\Big{(}\sqrt{p_u\kappa_{k,t}}\textbf{P}_{k}^{1/2}\textbf{s}_{k}+\boldsymbol{\eta}_{k,t}^H\Big{)}\displaystyle+\textbf{V}_k^H\boldsymbol{\eta}_r+\textbf{V}_k^H\textbf{n}\\ &\displaystyle=\textbf{V}_k^H\hat{\textbf{G}}_k\sqrt{p_u\kappa_{k,t}}\textbf{P}_{k}^{1/2}\textbf{s}_{k}+\textbf{V}_k^H\tilde{\textbf{G}}_k\sqrt{p_u\kappa_{k,t}}\textbf{P}_{k}^{1/2}\textbf{s}_{k}\\&\displaystyle+\sum\limits_{k'\neq k}^{K}\textbf{V}_k^H\textbf{G}_{k'}\sqrt{p_u\kappa_{k',t}}\textbf{P}_{k'}^{1/2}\textbf{s}_{k'}+ 

\sum\limits_{k=1}^{K}\textbf{V}_k^H\textbf{G}_k\boldsymbol{\eta}_{k,t}^H

\\&\displaystyle+\textbf{V}_k^H\boldsymbol{\eta}_r+\textbf{V}_k^H\textbf{n}
,
\end{array}\label{s_detection_CPU}
   \end{equation}
where $\tilde{\textbf{G}}_{k}={\textbf{G}}_{k}-\hat{\textbf{G}}_{k}$ denotes for the collective channel estimation error. Based on \eqref{s_detection_CPU}, we apply the MMSE estimator to compute channel estimates for all users. Consequently, we can derive the achievable SE for the $k$-th user employing MMSE-SIC detectors and applying the standard capacity lower bounds \cite{9737367,8845768} as
\begin{equation}
\begin{array}{ll}
     \displaystyle \text{SE}_k=\frac{\tau_c-\tau_p}{\tau_c}\mathbb{E}\Bigg{\{}\text{log}_2\Big{|}\textbf{I}_{N_u}+\hat{\textbf{D}}_{{k}}^H\hat{\boldsymbol{\Sigma}}_k^{-1}\hat{\textbf{D}}_{{k}}\Big{|}\Bigg{\}},
\end{array}\label{uplink_SE_CPU}
   \end{equation}
where 
\begin{equation}
\begin{array}{ll}
     \displaystyle \hat{\textbf{D}}_{{k}}=\textbf{V}_k^H\hat{\textbf{G}}_k\sqrt{p_u\kappa_{k,t}}\textbf{P}_{k}^{1/2},
\end{array}\label{Dk_CPU}
   \end{equation}
\begin{equation}
\begin{array}{ll}
     \displaystyle \hat{\boldsymbol{\Sigma}}_k=\textbf{V}_k^H\left[\begin{array}{ll}
   \displaystyle\sum\nolimits_{k'=1}^Kp_u\hat{\textbf{G}}_{k'}\textbf{P}_{k'}\hat{\textbf{G}}_{k'}^H-p_u\kappa_{k,t}\hat{\textbf{G}}_{k}\textbf{P}_{k}\hat{\textbf{G}}_{k}^H\vspace{4 pt}\\+\displaystyle\sum\nolimits_{k'=1}^Kp_u\tilde{\textbf{C}}_{k'}+\textbf{C}_r+\sigma^2\textbf{I}_{MN_{ap}}     \end{array}
     \right]\textbf{V}_k,
\end{array}\label{Sigmak_CPU}
   \end{equation}
with
\begin{equation}
\begin{array}{ll}
     \displaystyle 
\tilde{\textbf{C}}_{k}=\text{blkdiag}\left(\kappa_{1,r}\tilde{\textbf{C}}_{1k},...,\kappa_{M,r}\tilde{\textbf{C}}_{Mk}\right),
\end{array}
   \end{equation}
  \begin{equation}
\begin{array}{ll}
     \displaystyle 
{\textbf{C}}_{r}=\text{blkdiag}\left(\bar{\textbf{C}}_{1|{\textbf{G}_{mk}}},...,\bar{\textbf{C}}_{M|{\textbf{G}_{mk}}}\right).
\end{array}
   \end{equation} 

Notably, any combining matrix $\textbf{V}_k$ can be applied in \eqref{s_detection_CPU}, and the Monte Carlo method is utilized to compute the achievable SE\cite{8845768,9737367}. The CPU can leverage all channel estimates to formulate $\textbf{V}_k$. Similar to Level 1, we introduce two combining schemes for Level 2, including MR combining $\textbf{V}_k=\left[
\hat{\textbf{G}}_{1k}^T,...,\hat{\textbf{G}}_{Mk}^T
\right]^T$, and the global MMSE combining to minimize $\text{MSE}_{k}=\mathbb{E}\{||\textbf{s}_k- {\textbf{V}}_{k}^H\textbf{y}||^2|\left[
\hat{\textbf{G}}_{1k}^T,...,\hat{\textbf{G}}_{Mk}^T
\right]^T\}$. The global MMSE combining is given by 
   \begin{equation}
\begin{array}{ll}
     \displaystyle {\textbf{V}}_{k}=\left[
     \begin{array}{ll}
    \displaystyle\sum\nolimits_{k'=1}^K p_u\left(\hat{\textbf{G}}_{k'}\textbf{P}_{k'}\hat{\textbf{G}}_{k'}^H+\tilde{\textbf{C}}_{k'}\right)\vspace{4 pt}\\
\displaystyle
     +\bar{\textbf{C}}_r+\sigma^2\textbf{I}_{MN_{ap}}      
     \end{array}\right]^{-1}\sqrt{p_u\kappa_{k,t}}\hat{\textbf{G}}_{k}\textbf{P}_k^{1/2}.
\end{array}\label{CPU_MMSE}
   \end{equation}
   \textit{Proof:} The Proof of \eqref{CPU_MMSE} follows a similar process to that of \eqref{Local_MMSE} and is therefore omitted for brevity.
   
   \textit{Remark 2:} It shows that the MMSE combining is designed to be the optimal combining matrix, effectively minimizing MSE and maximizing achievable SE, as indicated by \eqref{uplink_SE_CPU_optimal} at the top of the next page. Although the global MMSE combining involves higher computational complexity than Level 1, which can be compensated since Level 2 is implemented at the CPU with high computational capability.
\begin{figure*}[t!]
   \begin{equation}
\begin{array}{ll}
     \displaystyle \text{SE}_k^{\text{max}}=\frac{\tau_c-\tau_p}{\tau_c}\mathbb{E}\Bigg{\{}\text{log}_2\Bigg{|}\textbf{I}_{N_u}+p_u\kappa_{k,t}\textbf{P}_k^{1/2}\hat{\textbf{G}}_{k}^H\left[
\sum\limits_{k'=1}^Kp_u\hat{\textbf{G}}_{k'}\textbf{P}_{k'}\hat{\textbf{G}}_{k'}^H-p_u\kappa_{k,t}\hat{\textbf{G}}_{k}\textbf{P}_{k}\hat{\textbf{G}}_{k}^H+\sum\limits_{k'=1}^Kp_u\tilde{\textbf{C}}_{k'}+\textbf{C}_r+\sigma^2\textbf{I}_{MN_{ap}}     
     \right]^{-1}\hat{\textbf{G}}_{k}\textbf{P}_k^{1/2}\Bigg{|}\Bigg{\}}.
\end{array}\label{uplink_SE_CPU_optimal}
   \end{equation}
   \vspace{-10pt}
   \hrulefill
   \end{figure*}

\section{Numerical Results and Discussions}

This section presents numerical results that include MC simulations and closed-form analytical results. In particular, \eqref{uplink_SE_level1}-\eqref{Sigma_k_CF} determine the uplink SE performance for Level 1, and \eqref{uplink_SE_CPU}-\eqref{uplink_SE_CPU_optimal} deliver the uplink SE performance for Level 2.
Our primary focus is demonstrating the performance analysis of the STAR-RIS-assisted cell-free massive MIMO system with multi-antenna users and exploring the impact of hardware impairments and various signal processing implementations with arbitrary combining schemes.

\subsection{Parameter Setup}

A two-dimensional coordinate system akin to \cite{qian2024performance,10297571} is applied in this section. APs are randomly distributed around the origin with $x^{\text{AP}},y^{\text{AP}}\in\left[-100,100\right]$. Users in the reflection area are located with $x^{\text{user}}\in\left[400,600\right]$
and $y^{\text{user}}\in\left[0,100\right)$, while users in the transmission area are distributed with $x^{\text{user}}\in\left[400,600\right]$
and $y^{\text{user}}\in(100,200]$. The STAR-RIS is positioned at $(x^{\text{STAR-RIS}},y^{\text{STAR-RIS}})=(500,100)$. All the geographic settings are in meter units. The AP, user and RIS heights are 15m, 1.65m, and 30m, respectively \cite{10225319}. 
Our path loss model, as introduced in \cite{qian2024performance}, employs relevant settings to express large-scale fading coefficients as $\beta_x=\text{PL}_x\cdot z_x$ 
($x=mk,~m,~k$). $\text{PL}_x$ is the three-slope path loss, $z_x$ is the log-normal shadowing. Unless otherwise stated, $\kappa_{m,r}=\kappa_{ap}=0.9,~\forall m$, $\kappa_{k,t}=\kappa_{u}=0.95,~\forall k$, $p_{p}=p_u=20~\text{dBm}$, $p_{d}=23~\text{dBm}$, $\sigma^2=-91$ dBm, $\xi_{kn}=1/N_u,~\forall k,~\forall m$. The AP spatial correlation model follows the exponential correlation model as introduced in \cite{951380}, $d_k=\lambda/4,~\forall k,$ for all mutli-antenna users and $d_h=d_v=\lambda/4$ for the STAR-RIS elements. Additionally, each coherence block comprises $\tau_c=200$ symbols, $\tau_p=KN_u/2$.

\subsection{Effects of the Number of Antennas Per User}
\begin{figure}[!t]
\centering
\includegraphics[width=0.84\columnwidth,height=0.63\columnwidth]{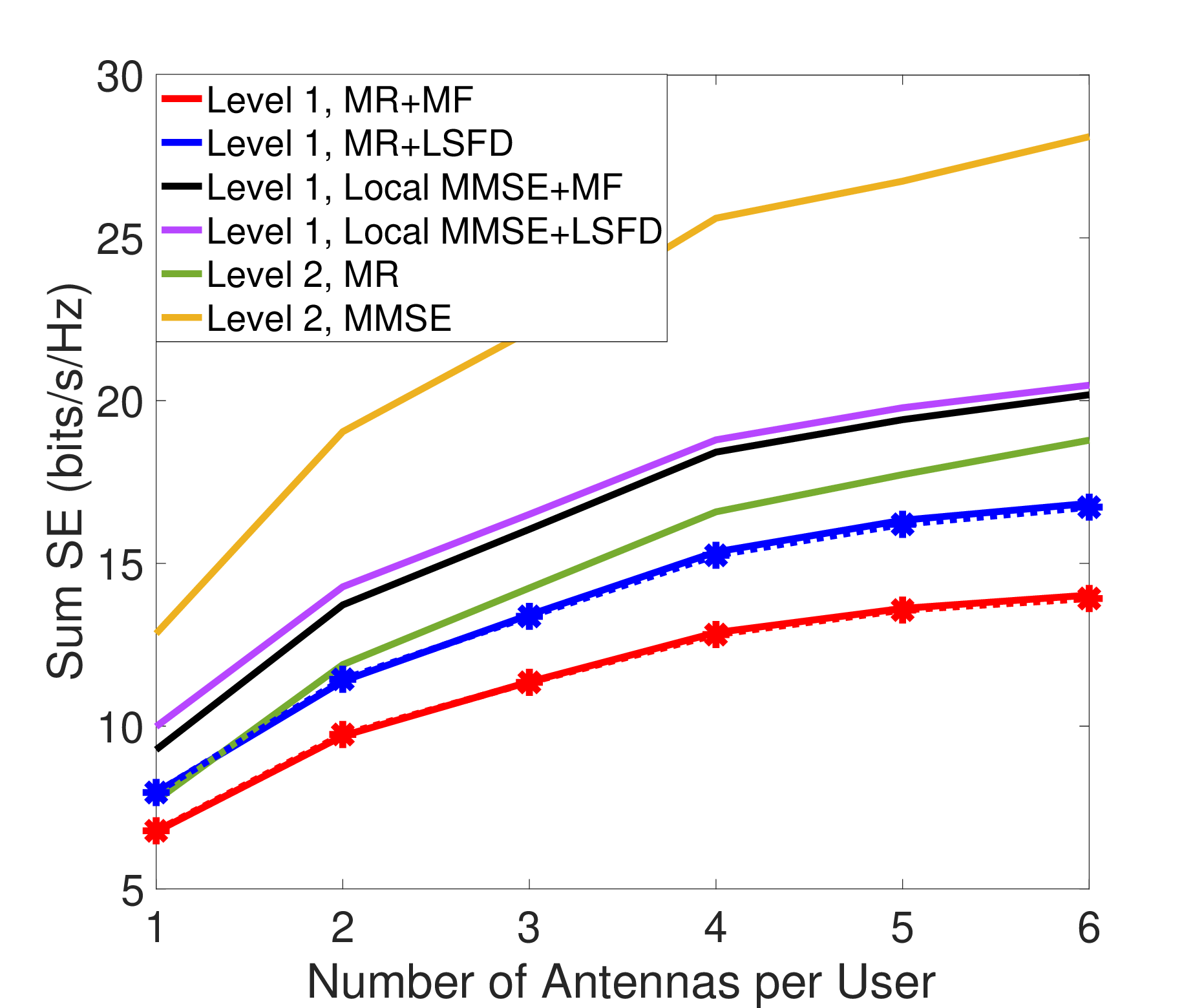}
\caption{Sum SE v.s. Number of Antennas per User with $M=20$, $K=10$, $N_{ap}=4$, $L=16$. (MC Simulations: Solid lines; Analytical Results: Dashed lines with markers)}
\label{fig_1}
\vspace{-5 pt}
\end{figure}
\begin{figure}[!t]
\centering
\includegraphics[width=0.84\columnwidth,height=0.63\columnwidth]{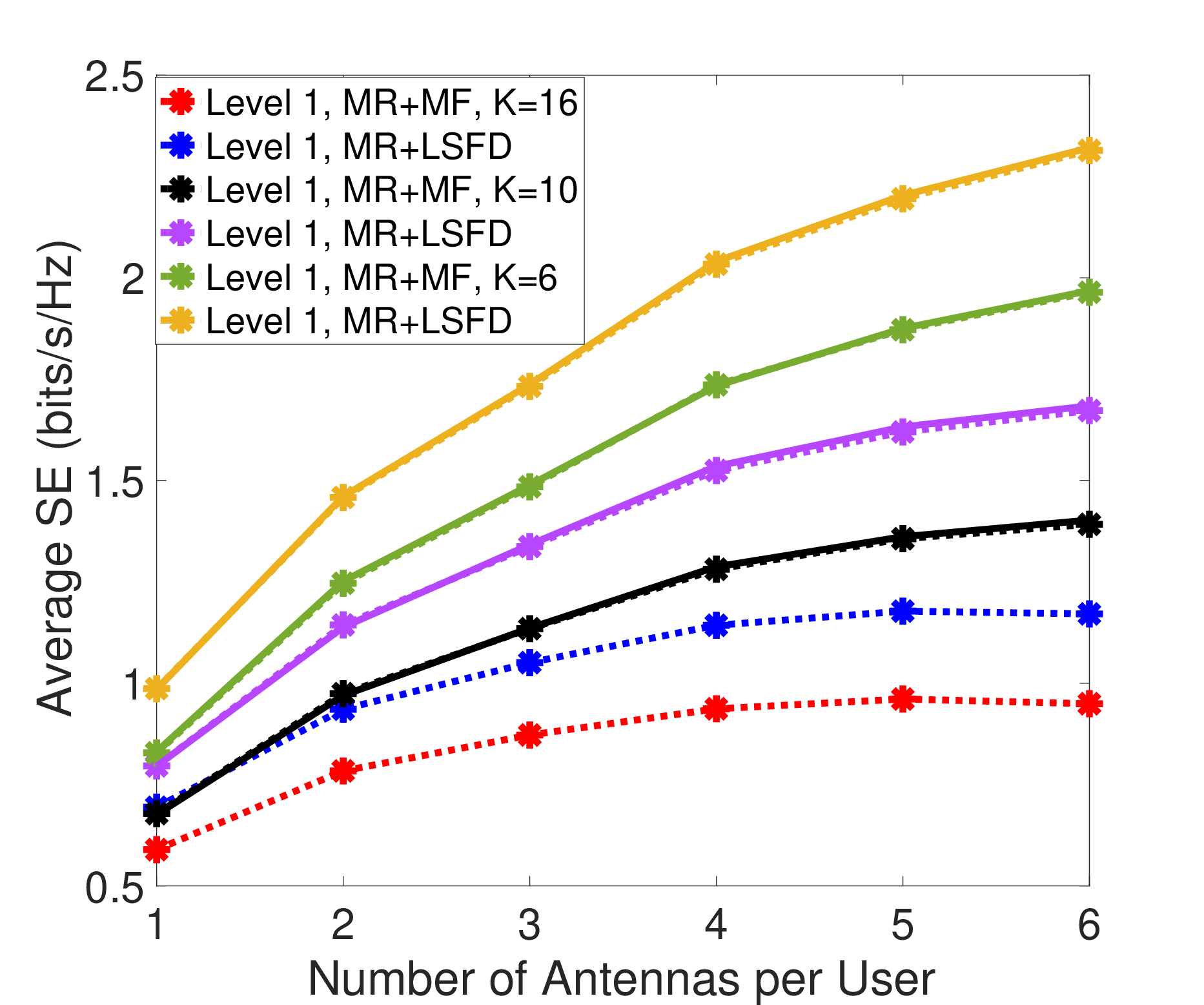}
\caption{Average SE v.s. Number of Antennas per User with $M=20$, $N_{ap}=4$, $L=16$. (MC Simulations: Solid lines; Analytical Results: Dashed lines with markers).}
\label{fig_2}
\vspace{-4 pt}
\end{figure}
Fig. \ref{fig_1} illustrates the sum SE, denoted as $\sum\nolimits_{k=1}^K\text{SE}_k$, as a function of the number of antennas per user, $N_u$. The proposed two processing implementations utilizing MMSE combining and MR combining are considered. The closed-form analytical results based on \eqref{uplink_SE}-\eqref{Sigma_k_CF} for Level 1 can closely match the numerical results obtained from MC simulations. The results reveal that increasing the number of antennas per user can greatly improve the system performance. However, the growth trend of SE performance slows as the number of antennas per user increases. This indicates that increasing the number of antennas per user leads to increasing interference and hardware impairments, leading to a reduction of SE improvement. 
{\color{black} A significant finding from our results is that Level 2 can substantially outperform Level 1 using the same combining scheme, especially with a larger number of user antennas, primarily due to the introduction of the STAR-RIS, which enhances overall performance. This contrasts with conventional direct link-only results, where Level 1 with LSFD decoding exhibits superior performance compared to Level 2 when MR combining is applied\cite{9737367}.} 
The performance gap between Level 1 with MF decoding and LSFD decoding using MR combining is larger than that of local MMSE combining. It shows that utilizing Level 1 with local MMSE and Level 2 can enhance the performance and offset the effects of hardware impairments.
Increasing the number of antennas per user reduces the pre-log factor $(\tau_c - \tau_p) / \tau_c$, which outweighs the gain from having more user antennas. {\color{black}{Please note that this work employs a simple equal power allocation strategy. However, implementing efficient power allocation optimization could further enhance the advantages of incorporating multi-antenna users. This aspect will be addressed in our future research.}}


\subsection{Effects of the Number of Users}

In Figure \ref{fig_2}, the average SE per user, denoted as $\frac{1}{K}\sum\nolimits_{k=1}^K\text{SE}_k$, is plotted against the number of antennas per user $N_u$ for Level 1 with MR combining. For $K=6$ and $K=10$, it is observed that $N_u=6$
can achieve respective $130\%$ and $110\%$-likely improvement in average SE compared with $N_u=1$ when applying LSFD decoding, respectively. This result reveals that, in scenarios with a small or moderate number of users, the average SE performance benefits from increasing the number of antennas per user to increase spatial multiplexing.
However, the benefits of introducing more antennas per user on the average SE may decrease significantly with numerous users (such as $K=16$). 
Moreover, with the increasing number of users, the average performance diminishes. For example, $K=6$ achieves a $100\%$-likely larger average SE than $K=16$ when $N_u=6$. This degradation is attributed to the higher inter-user interference and pilot contamination introduced by more users. {\color{black}Thus, it is crucial to introduce advanced channel estimation schemes or interference elimination schemes, like learning-based channel estimation \cite{10449720}, to improve system performance in future work.}

\subsection{Effects of the Hardware Impairment Levels}
\begin{figure}[!t]
\centering
\includegraphics[width=0.84\columnwidth,height=0.63\columnwidth]{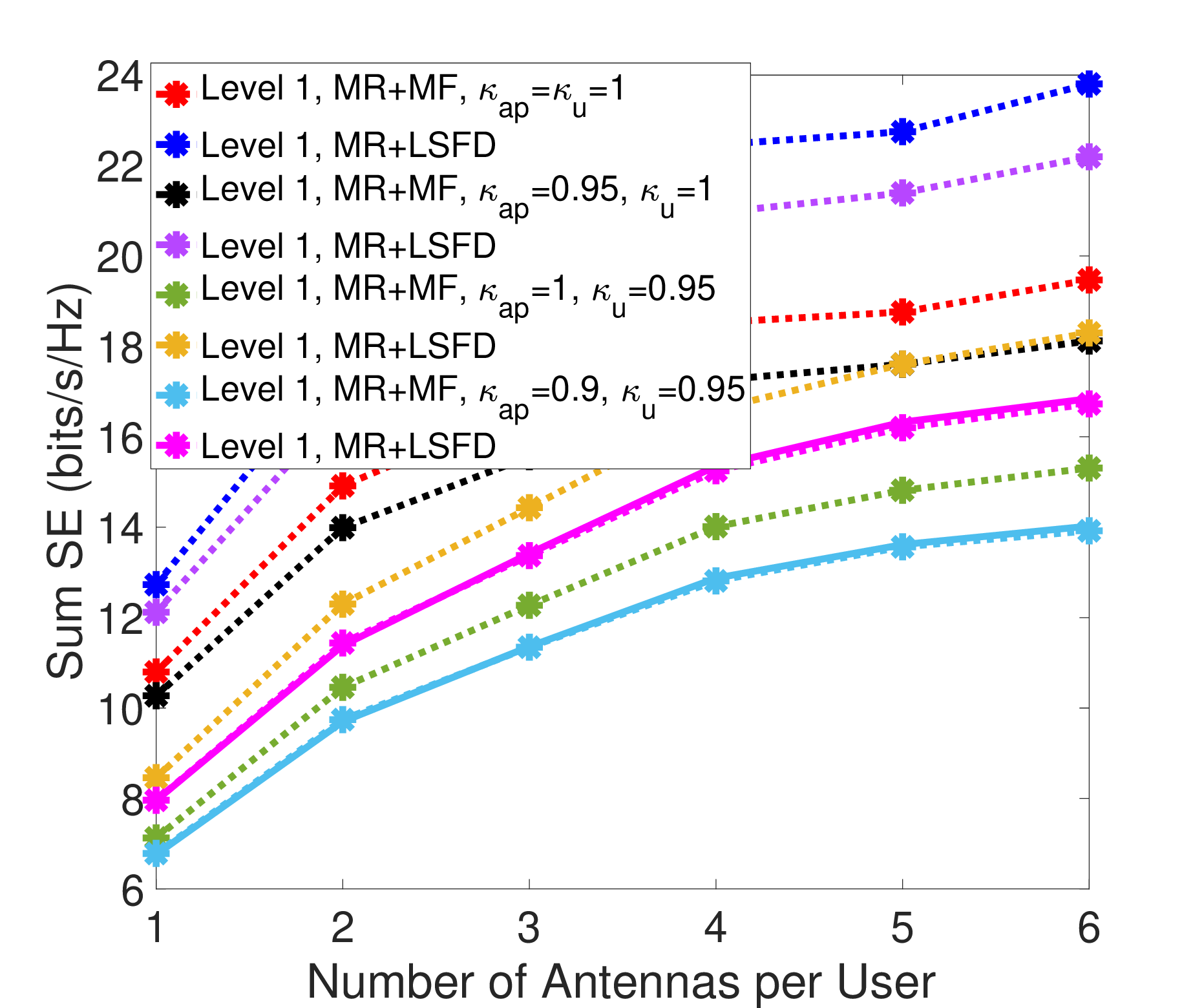}
\caption{Sum SE v.s. Number of Antennas per User with $M=20$, $K=10$, $N_{ap}=4$, $N_{u}=4$, $L=16$. (MC Simulations: Solid lines; Analytical Results: Dashed lines with markers).}
\label{fig_3}
\vspace{-12 pt}
\end{figure}
In fig. \ref{fig_3}, we illustrate the impact of the different hardware qualities on the SE performance. 
When both APs and users are experiencing hardware impairments with $\kappa_{ap}, ~\kappa_u \neq 1 $, the SE performance experiences significant performance degradation. For example, APs and users with ideal hardware components, namely, $\kappa_{ap}=\kappa_u = 1 $, can achieve a nearly $60\%$-likely larger SE than the scenario with $\kappa_{ap}=0.9,~ \kappa_u = 0.95 $ when $N_u=6$.  
Moreover, the scenario with $\kappa_{ap}=0.95,~\kappa_{u}=1 $ achieves an around $20\%$-likely larger SE performance compared with $\kappa_{ap}=1,~\kappa_{u}=0.95 $ when $N_u=6$. This result reveals that the hardware impairments experienced by users introduce a higher performance degradation than that of APs. Moreover, it shows that more user antennas, e.g., $N_u=6$, can achieve more than $80\%$-likely larger SE improvement than $N_u=1$ under given scenarios. This result implies that introducing more user antennas can offset the performance degradation caused by hardware impairments, which motivates the deployment of multi-antenna users. Additionally, the results indicate that the impact of hardware impairments is non-negligible in practical scenarios and underscore the necessity to introduce hardware-impairment elimination methods to improve the system performance in practical scenarios.

\subsection{Effects of the Number of APs and Antennas Per AP}
\begin{figure}[!t]
\centering
\includegraphics[width=0.84\columnwidth,height=0.63\columnwidth]{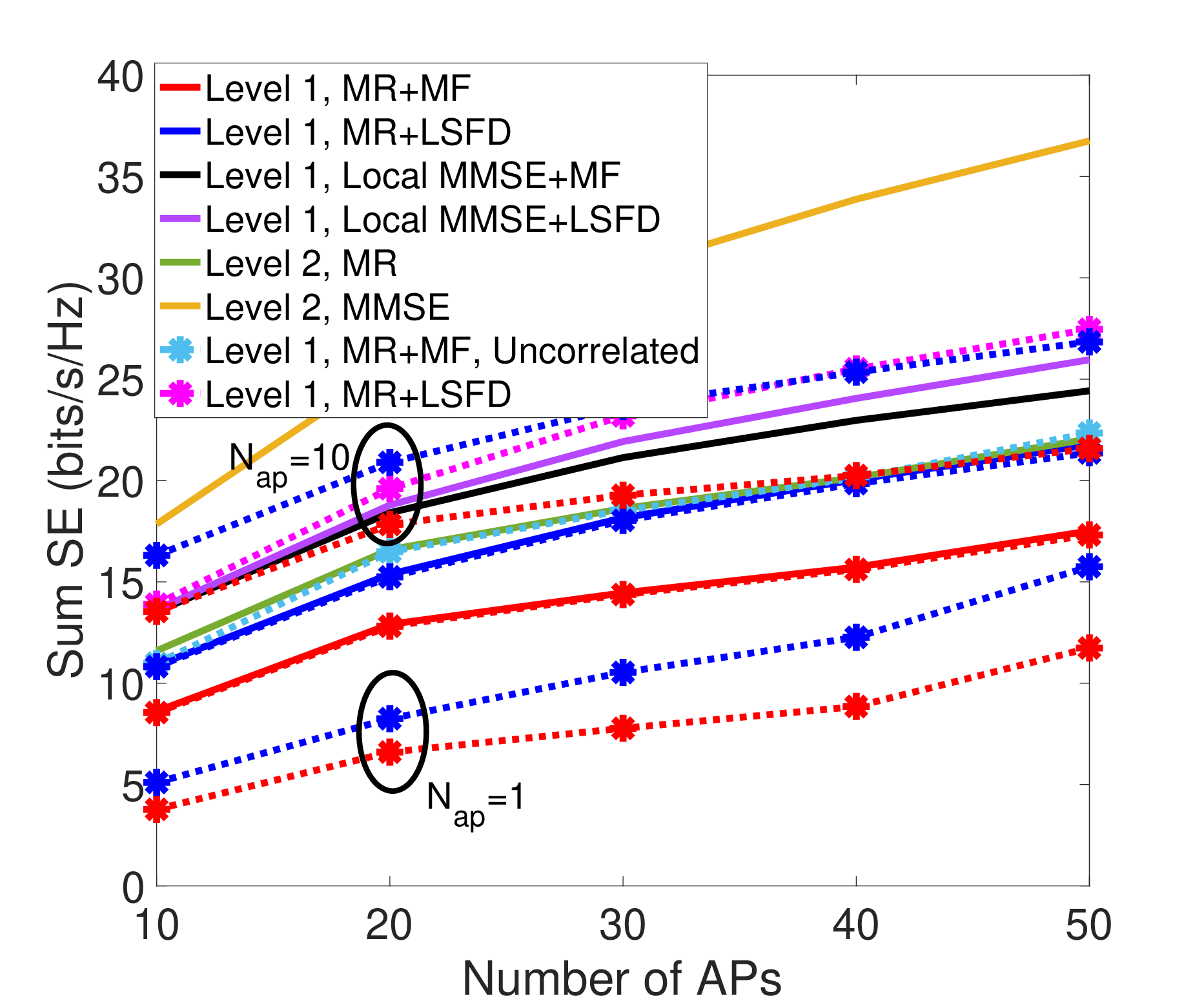}
\caption{Sum SE v.s. Number of APs with $K=10$, $N_{ap}=4$ (until mentioned), $N_{u}=4$, $L=16$. (MC Simulations: Solid lines; Analytical Results: Dashed lines with markers).}
\label{fig_4}
\vspace{-6 pt}
\end{figure}
Fig. \eqref{fig_4} displays the sum SE as a function of the number of APs, with varying numbers of antennas per AP. Undoubtedly, the
sum SE increases with the number of APs and antennas per AP. For example, $M=50$ can introduce an around $100\%$-likely larger SE than $M=10$ with $N_{ap}=4$. Meanwhile, $N_{ap}=10$ offers a $150\%$-likely SE improvement than single-antenna APs at $M=20$. This indicates that increasing APs and antennas per AP can introduce higher spatial freedom to enhance the sum SE and offset the system degradation caused by hardware impairments and spatial correlations. However, the growth trend of performance gains diminishes with the increasing number of APs and antennas per AP due to heightened inter-user interference and network overhead. To this end, it is essential to strategically increase the number of APs and antennas per AP to achieve the desired performance. Notably, a larger number of APs will make the performance of Level 1 with LSFD decoding gradually approach that of Level 2 when MR combining is applied.
Meanwhile, Level 1, with uncorrelated APs and users, can achieve a $30\%$-likely higher SE than the correlated Rayleigh fading scenario, highlighting the impact of spatial correlation experienced by APs and users on channel hardening and SE performance. {\color{black}This underscores the importance of accounting for the spatially correlated Rayleigh fading channel, as the adverse effects of spatial correlation are significant due to the multi-antenna structure. It is worth noting that advanced beamforming techniques, such as MMSE combining, can help mitigate these negative effects. In our future work, we plan to explore additional advanced methods to either reduce the impact of spatial correlation or leverage it for improved performance\cite{8388873}.}

\subsection{Effects of the Number of STAR-RIS Elements}
\begin{figure}[t!]
    \centering
    \subfigure[]{\includegraphics[width=0.84\columnwidth,height=0.38\columnwidth]{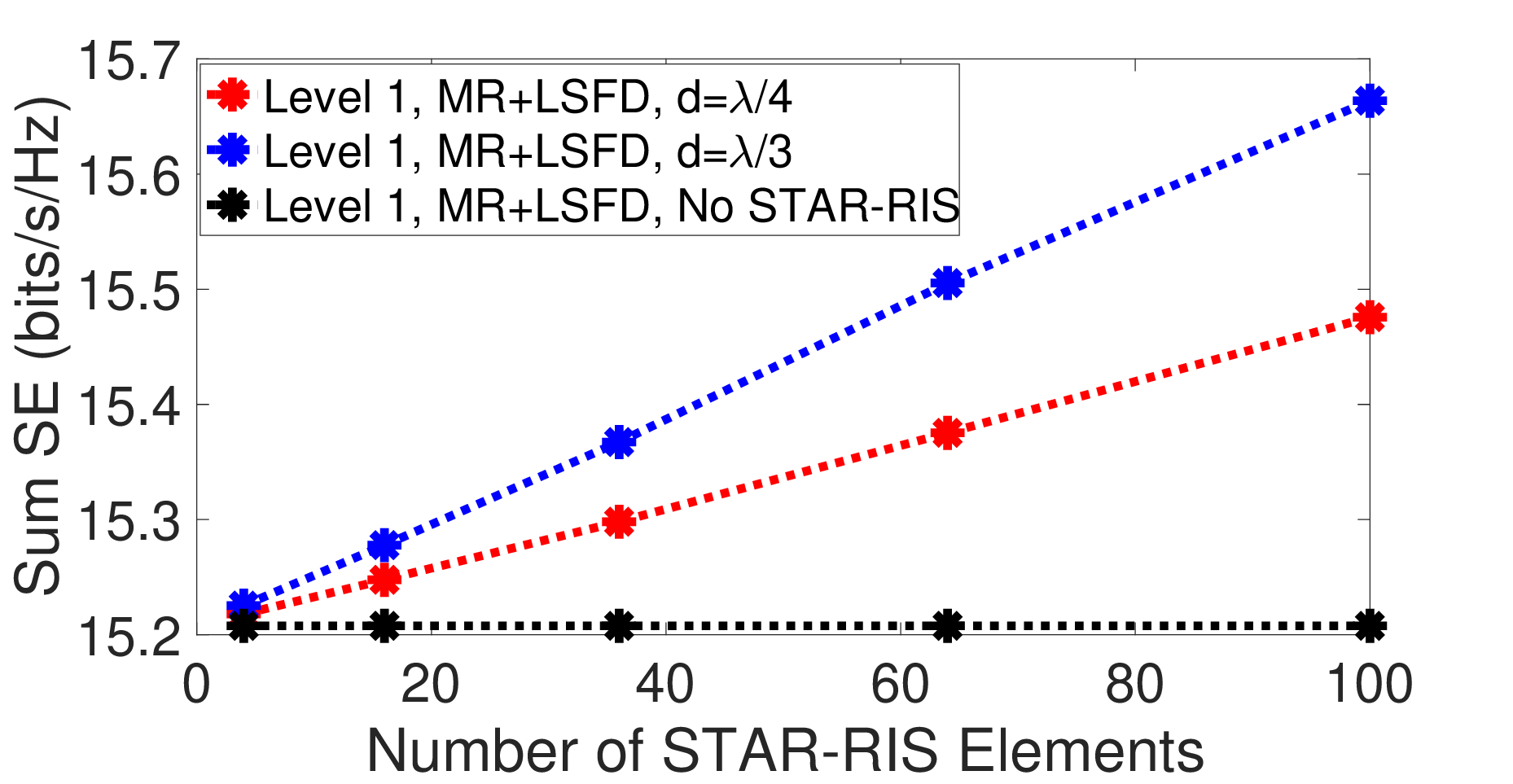}} 
    \vspace{-4 pt}
    \subfigure[]{\includegraphics[width=0.84\columnwidth,height=0.63\columnwidth]{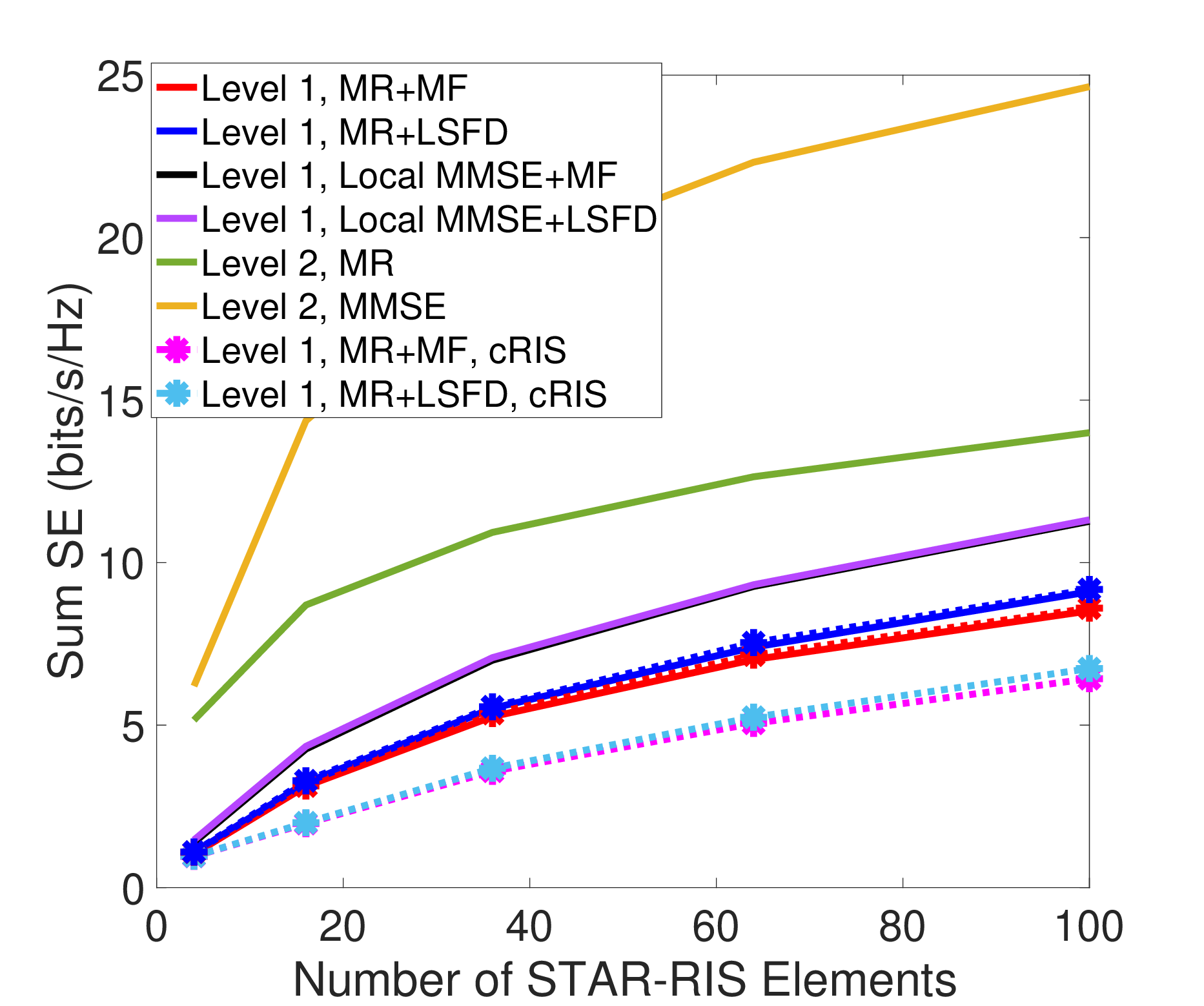}} 
    \caption{(a) Sum SE v.s. Number of STAR-RIS Elements with direct links and (b) Sum SE v.s. Number of STAR-RIS Elements without direct links. $M=20$, $K=10$, $N_{ap}=4$, $N_{u}=4$. (MC Simulations: Solid lines; Analytical Results: Dashed lines with markers).}
    \label{fig_6}
    \vspace{-4 pt}
\end{figure}

Fig. \ref{fig_6} introduces the sum SE as a function of the number of STAR-RIS elements. {\color{black}The conventional cell-free massive MIMO system without STAR-RISs is considered for comparison in Fig. 6 (a). It shows that the proposed system enhanced by a STAR-RIS with $d_h=d_v=\lambda/4$ can outperform conventional cell-free massive MIMO by $3\%$-likely SE improvement when $L=100$. Meanwhile, a larger STAR-RIS element distance, $d_h=d_v=\lambda/3$, results in additional performance gains. Therefore, introducing STAR-RISs can improve performance and offset the negative effects of transceiver hardware impairments. Additionally, we study the SE performance when direct links are blocked in Fig. 6 (b) to highlight the benefits of introducing more STAR-RIS elements. It reveals that increasing the number of STAR-RIS elements can greatly improve system performance.} 
We also introduce a benchmark scenario involving a conventional reflecting-only RIS close to a transmitting-only RIS positioned at the STAR-RIS location. In fairness, it is assumed that each conventional RIS (cRIS) consists of $L/2$ elements \cite{9570143,qian2024performance}. The results indicate that the proposed STAR-RIS-assisted system can introduce a $30\%$-likely SE improvement compared to cRISs, highlighting the benefits of integrating STAR-RISs into cell-free massive MIMO systems. Compared to Fig. \ref{fig_1}, Fig. \ref{fig_6} introduces another remarkable insight: Level 2 outperforms Level 1 by a large margin when direct links are blocked, regardless of the combining schemes. 
Specifically, Level 2 with MR combining introduces more than $20\%$-likely SE improvement over Level 1, and Level 2 with MMSE combining achieves more than $110\%$-likely SE improvement over Level 1.
Thus, when direct links are indistinctive or blocked, Level 2 introduces greater advancement in STAR-RIS-assisted cell-free massive MIMO systems.

\section{Conclusion}
STAR-RIS-assisted cell-free massive MIMO has garnered significant research interest due to its sustainable architecture that leverages their combined advantages. To the best of our knowledge, this is the first study to incorporate multi-antenna users in spatially correlated STAR-RIS-assisted cell-free massive MIMO systems, considering the transceiver hardware impairments. With this model, the SE performance analysis has been investigated. Two uplink implementations with arbitrary combining schemes are studied. The uplink MMSE combining scheme, which can maximize the system performance, is developed. Moreover, we explored the LSFD decoding for local processing and centralized decoding.
We also derived the closed-form uplink SE expressions for local processing and centralized decoding with the MR combining scheme for performance evaluation. It shows that hardware impairments cause non-negligible degradation in sum SE, particularly at the user end. Increasing the number of user antennas offsets the system degradation caused by hardware impairments. Specifically, users with six antennas can introduce over $100\%$-likely SE improvement compared to single-antenna users. Thus, it is beneficial to employ multi-antenna users in future wireless networks. Also, increasing APs and STAR-RIS elements can reduce performance degradation. Note that introducing STAR-RIS can achieve a $3\%$-likely larger SE than conventional cell-free massive MIMO, and STAR-RIS can achieve a $30\%$-likely larger SE than conventional RISs when direct links are blocked, introducing the necessity of deploying STAR-RISs to improve performance and offset
the negative effects of transceiver hardware impairments.
Remarkably, distinct from existing results, fully centralized processing consistently outperforms local processing and centralized decoding in STAR-RIS-assisted cell-free massive MIMO systems when direct links are indistinctive or blocked, indicating the benefits of fully centralized processing for performance improvement. The above-mentioned results motivate our future work, focusing on the advanced inter-user interference mitigation method and hardware impairment elimination method to leverage the advantages of deploying STAR-RISs and serving multi-antenna users sufficiently.

\ifCLASSOPTIONcaptionsoff
  \newpage
\fi

\begin{appendices}
\section
{Derivation of MMSE Channel Estimation}
 \label{Appendix_NMSE}
First, the MMSE channel estimation is defined as \cite{9459571}
 \begin{equation}
\begin{array}{ll}
     \displaystyle \hat{\textbf{g}}_{mk}=\frac{\mathbb{E}\{\mathbf{g}_{mk}\mathbf{y}_{mk,p}^H\}}{\mathbb{E}\{\mathbf{y}_{mk}\mathbf{y}_{mk,p}^H\}}\mathbf{y}_{mk,p}.
\end{array}
   \end{equation}

According to Kronecker product and vectorization, we can calculate $\mathbf{Q}_{mk}=\mathbb{E}\{\mathbf{g}_{mk}\mathbf{y}_{mk,p}^\text{H}\}$. Since the channel response of the $k$-th user is uncorrelated with the channel vector of $k'$-th user, $\forall k' \neq k$, hardware impairments and noise, it yields
 \begin{equation}
\begin{array}{ll}
     \displaystyle \mathbf{Q}_{mk}\displaystyle=\mathbb{E}\left\{ {\textbf{g}}_{mk}\textbf{y}_{mk,p}^H\right\}\\ ~~~~~\displaystyle=\sqrt{\kappa_{m,r}}\sum\limits_{k'\in\mathcal{P}_k}\sqrt{\tau_pp_{p}\kappa_{k',t}}\mathbb{E}\left\{ \textbf{g}_{mk}\textbf{g}_{mk'}^H\left(\textbf{P}_{k'}^{1/2}\otimes\textbf{I}_{N_{ap}}\right)^H\right\}
     \\ ~~~~~\displaystyle=\sqrt{\tau_pp_{p}\kappa_{m,r}\kappa_{k,t}}\bar{{\Delta}}_{mk}\Big{(}\textbf{R}_{k,t}\textbf{P}_{k}^{1/2}\otimes\textbf{R}_{m,r}\Big{)}.
\end{array}
   \end{equation}
Similarly, the covariance of $\mathbf{y}_{mk,p}$,  $\mathbf{\Psi}_{mk}=\mathbb{E}\Big{\{} {\textbf{y}}_{mk}\textbf{y}_{mk,p}^H\Big{\}}$ equals to \eqref{Psi_derivation} at the top of the next page.
\begin{figure*}[!t]
\begin{equation}
\begin{array}{ll}
     \displaystyle \mathbf{\Psi}_{mk}&\displaystyle=\mathbb{E}\{\mathbf{y}_{mk,p}\mathbf{y}_{mk,p}^H\}\\ &\displaystyle=\underbrace{\kappa_{m,r}\sum\limits_{k'\in\mathcal{P}_k}\tau_pp_p\kappa_{k',t}\mathbb{E}\left\{\left(\textbf{P}_{k'}^{1/2}\otimes\textbf{I}_{N_{ap}}\right)\textbf{g}_{mk'}\textbf{g}_{mk'}^H \left(\textbf{P}_{k'}^{1/2}\otimes\textbf{I}_{N_{ap}}\right)^H\right\}}_{\textbf{V}_{1mk}}+\underbrace{   \kappa_{m,r}\sum\limits_{k'=1}^K\mathbb{E}\left\{\left((\textbf{W}_{k',t}^{\text{H}}\mathbf{\Phi}_{k})^T\otimes\textbf{I}_{N_{ap}}\right)\textbf{g}_{mk'}\textbf{g}_{mk'}^H \left((\textbf{W}_{k',t}^{\text{H}}\mathbf{\Phi}_{k})^T\otimes\textbf{I}_{N_{ap}}\right)^H\right\}}_{\textbf{V}_{2mk}}\\&\displaystyle+\underbrace{  \mathbb{E}\left\{\left(\mathbf{\Phi}_{k}^T\otimes\textbf{I}_{N_{ap}}\right)\text{vec}(\textbf{W}_{m,r})\text{vec}(\textbf{W}_{m,r})^H \left(\mathbf{\Phi}_{k}^T\otimes\textbf{I}_{N_{ap}}\right)^H\right\}}_{\textbf{V}_{3mk}}+\underbrace{\mathbb{E}\left\{\left(\mathbf{\Phi}_{k}^T\otimes\textbf{I}_{N_{ap}}\right)\text{vec}(\textbf{N}_{m,p})\text{vec}(\textbf{N}_{m,p})^H \left(\mathbf{\Phi}_{k}^T\otimes\textbf{I}_{N_{ap}}\right)^H\right\}}_{\textbf{V}_{4mk}}.  
\end{array}
\label{Psi_derivation}
   \end{equation}
   \vspace{-10 pt}
   \hrulefill
\end{figure*}
Then, $\mathbf{\Psi}_{mk}$ can be decomposed into four parts with $\textbf{V}_{1mk}$ and $\textbf{V}_{2mk}$ are given by \eqref{V_1mk}-\eqref{V_2mk} at the top of the next page, while $\textbf{V}_{3mk}$ and $\textbf{V}_{4mk}$ are given by
\begin{figure*}
\begin{equation}
\begin{array}{ll}
     \displaystyle \textbf{V}_{1mk}&\displaystyle=\kappa_{m,r}\sum\limits_{k'\in\mathcal{P}_k}\tau_pp_p\kappa_{k',t}\mathbb{E}\left\{\left(\textbf{P}_{k'}^{1/2}\otimes\textbf{I}_{N_{ap}}\right)\textbf{g}_{mk'}\textbf{g}_{mk'}^H \left(\textbf{P}_{k'}^{1/2}\otimes\textbf{I}_{N_{ap}}\right)^H\right\}\\&\displaystyle=\kappa_{m,r}\sum\limits_{k'\in\mathcal{P}_k}\tau_pp_p\kappa_{k',t}\left(\textbf{P}_{k'}^{1/2}\otimes\textbf{I}_{N_{ap}}\right)\mathbf{\Delta}_{mk'} \left(\textbf{P}_{k'}^{1/2}\otimes\textbf{I}_{N_{ap}}\right)^H\\&\displaystyle=\kappa_{m,r}\sum\limits_{k'\in\mathcal{P}_k}\tau_pp_p\kappa_{k',t}\Bigg{(}\beta_{mk'}\left(\textbf{P}_{k'}^{1/2}\otimes\textbf{I}_{N_{ap}}\right)\left({\textbf{R}}_{k',t} \otimes{\textbf{R}}_{m,r}\right)\left(\textbf{P}_{k'}^{1/2}\otimes\textbf{I}_{N_{ap}}\right)^H+\beta_{m}\beta_{k'}\text{tr}(\textbf{T}_{\omega_{k'}})\left(\textbf{P}_{k'}^{1/2}\otimes\textbf{I}_{N_{ap}}\right)\left({\textbf{R}}_{k',t} \otimes{\textbf{R}}_{m,r}\right)\left(\textbf{P}_{k'}^{1/2}\otimes\textbf{I}_{N_{ap}}\right)^H\Bigg{)}\\&\displaystyle=\kappa_{m,r}\sum\limits_{k'\in\mathcal{P}_k}\tau_pp_{p}\kappa_{k',t}\bar{{\Delta}}_{mk'}\bigg{(}\Big{(}\textbf{P}_{k'}^{1/2}\textbf{R}_{k',t}\textbf{P}_{k'}^{1/2}\Big{)}\otimes\textbf{R}_{m,r}\bigg{)},
\end{array}
   \label{V_1mk}
   \end{equation}
   \vspace{-10 pt}
  \hrulefill
\end{figure*}
\begin{figure*}
\begin{equation}
\begin{array}{ll}
     \displaystyle \textbf{V}_{2mk}&\displaystyle=\kappa_{m,r}\sum\limits_{k'=1}^K\mathbb{E}\left\{\left((\textbf{W}_{k',t}^{H}\mathbf{\Phi}_{k})^T\otimes\textbf{I}_{N_{ap}}\right)\textbf{g}_{mk'}\textbf{g}_{mk'}^H \left((\textbf{W}_{k',t}^{H}\mathbf{\Phi}_{k})^T\otimes\textbf{I}_{N_{ap}}\right)^H\right\}\\&\displaystyle=\kappa_{m,r}\sum\limits_{k'=1}^K\mathbb{E}\left\{\left((\textbf{W}_{k',t}^{H}\mathbf{\Phi}_{k})^T\otimes\textbf{I}_{N_{ap}}\right)\mathbf{\Delta}_{mk'} \left((\textbf{W}_{k',t}^{H}\mathbf{\Phi}_{k})^T\otimes\textbf{I}_{N_{ap}}\right)^H\right\}\\&\displaystyle=\kappa_{m,r}\sum\limits_{k'=1}^K\bar{{\Delta}}_{mk'}\mathbb{E}\left\{\left((\textbf{W}_{k',t}^{H}\mathbf{\Phi}_{k})^T\otimes\textbf{I}_{N_{ap}}\right)\left({\textbf{R}}_{k',t} \otimes{\textbf{R}}_{m,r}\right)\left((\textbf{W}_{k',t}^{H}\mathbf{\Phi}_{k})^T\otimes\textbf{I}_{N_{ap}}\right)^H\right\}\\&\displaystyle=\kappa_{m,r}\sum\limits_{k'=1}^K(1-\kappa_{k',t})p_p\bar{{\Delta}}_{mk'}\bigg{(}\text{tr}\Big{(}\textbf{P}_{k'}\textbf{R}_{k',t}\Big{)}\textbf{I}_{N_u}\otimes\textbf{R}_{m,r}\bigg{)},
\end{array}
\label{V_2mk}
   \end{equation}
  \hrulefill
\end{figure*}
\begin{equation}
    \begin{array}{ll}
    \displaystyle\textbf{V}_{3mk}&\displaystyle=\mathbb{E}\left\{\left(\mathbf{\Phi}_{k}^T\otimes\textbf{I}_{N_{ap}}\right)\text{vec}(\textbf{W}_{m,r})\text{vec}(\textbf{W}_{m,r})^H \left(\mathbf{\Phi}_{k}^T\otimes\textbf{I}_{N_{ap}}\right)^H\right\}\\&\displaystyle=\left(\mathbf{\Phi}_{k}^T\otimes\textbf{I}_{N_{ap}}\right)\left(\textbf{I}_{\tau_p}\otimes \textbf{C}_{m|\{\textbf{G}_{mk}\}}\right)\left(\mathbf{\Phi}_{k}^T\otimes\textbf{I}_{N_{ap}}\right)^H\\&\displaystyle=\left(\textbf{I}_{N_u}\otimes\textbf{C}_{m|\{\textbf{G}_{mk}\}}\right),
    \end{array}
\end{equation}
\begin{equation}
    \begin{array}{ll}
\displaystyle\textbf{V}_{4mk}&\displaystyle=\mathbb{E}\left\{\left(\mathbf{\Phi}_{k}^T\otimes\textbf{I}_{N_{ap}}\right)\text{vec}(\textbf{N}_{m,p})\text{vec}(\textbf{N}_{m,p})^H \left(\mathbf{\Phi}_{k}^T\otimes\textbf{I}_{N_{ap}}\right)^H\right\}\\&\displaystyle=\left(\mathbf{\Phi}_{k}^T\otimes\textbf{I}_{N_{ap}}\right)\left(\textbf{I}_{\tau_p}\otimes \sigma^2\textbf{I}_{N_{ap}}\right)\left(\mathbf{\Phi}_{k}^T\otimes\textbf{I}_{N_{ap}}\right)^H\\&\displaystyle=\sigma^2\left(\textbf{I}_{N_u}\otimes\textbf{I}_{N_{ap}}\right).
    \end{array}
\end{equation}

\section{Derivations of Uplink SE Expressions}
\label{Appendix_SE}
In this appendix, we present the detailed derivation of every term of $\text{SINR}_k$ by applying the use-and-then-forget bound \cite{8845768,9416909} to average the SE of $k$-th user. $\hat{\textbf{g}}_{mk'}$ is correlated with ${\textbf{g}}_{mk}$ for users sharing the same pilot sequence, namely, $k'\in\mathcal{P}_k$. 

\subsection{Compute Desired Signal} First, we can obtain the closed-form expression of desired signal $\bar{\textbf{D}}_k$ with $\bar{{\textbf{H}}}_{kk}= \mathbb{E}\Big{\{}\textbf{H}_{kk}\Big{\}}$, where
  \begin{equation}
\begin{array}{ll}
     \displaystyle \bar{{\textbf{H}}}_{kk}=[\bar{{\textbf{H}}}_{1kk}^T,...,\bar{{\textbf{H}}}_{Mkk}^T]^T\in\mathbb{C}^{MNu\times N_u},
\end{array}\label{D_k_CF1}
   \end{equation} 
with $\bar{{\textbf{H}}}_{mkk}\in\mathbb{C}^{N_u\times N_u}$, the $n,n'$-th element in $\bar{{\textbf{H}}}_{mkk}=\mathbb{E}\{\sqrt{\kappa_{m,r}\kappa_{k,t}}{\textbf{V}}_{mk}^H{\textbf{G}}_{mk}\}$, given by
\begin{equation}
\begin{array}{ll}
     \displaystyle [\bar{{\textbf{H}}}_{mkk}]_{n,n'}&\displaystyle=[\mathbb{E}\{\sqrt{\kappa_{m,r}\kappa_{k,t}}\hat{\textbf{G}}_{mk}^H{\textbf{G}}_{mk}\}]_{n,n'}\\ &\displaystyle=\sqrt{\kappa_{m,r}\kappa_{k,t}}[\mathbb{E}\{\hat{\textbf{G}}_{mk}^H\hat{\textbf{G}}_{mk}\}]_{n,n'}
     \displaystyle=\sqrt{\kappa_{m,r}\kappa_{k,t}}\text{tr}\left(\hat{\boldsymbol{\Delta}}_{mk}^{n,n'}\right),
\end{array}\label{D_k_CF2}
   \end{equation} 
where $\hat{\boldsymbol{\Delta}}_{mk}^{n,n'}\in\mathbb{C}^{N_{ap}\times N_{ap}}$ is the $n,n'$-th sub-matrix in $\hat{\boldsymbol{\Delta}}_{mk}$ with $\hat{\boldsymbol{\Delta}}_{mk}^{n,n'}=\hat{\boldsymbol{\Delta}}_{mk}\Big{(}(n-1)N_{ap}+1:nN_{ap},(n'-1)N_{ap}+1:n'N_{ap}\Big{)}$.

\subsection{Compute Inter-user Interference $\&$ Transmission Distortion}
We first introduce $ \textbf{U}_{kk'}\in\mathbb{C}^{MN_u\times M N_u}$, given by
\begin{equation}
    \begin{array}{ll}
    \textbf{U}_{kk'}=
\mathbb{E}\Big{\{}
\textbf{F}_{kk'}\textbf{P}_{k'}\textbf{F}_{kk'}^H\Big{\}},
        \end{array}
        \label{derivative_SE_2}
\end{equation}
where the $m,m'$-th sub-matrix,  $ \textbf{U}_{kk'}^{m,m'}\in\mathbb{C}^{N_u\times  N_u}$, is expressed as
\begin{equation}
    \begin{array}{ll}
    \textbf{U}_{kk'}^{m,m'}=\sqrt{\kappa_{m,r}\kappa_{m',r}}
\mathbb{E}\Big{\{}
\hat{\textbf{G}}_{mk}^H{\textbf{G}}_{mk'}{\textbf{P}}_{k'}{\textbf{G}}_{m'k'}^H\hat{\textbf{G}}_{m'k}\Big{\}}.
        \end{array}
        \label{derivative_SE_3}
\end{equation}
Then, \eqref{derivative_SE_3} can be calculated based on the following cases with $\textbf{Z}_{mk}=\textbf{Q}_{mk}\mathbf{\Psi}_{mk}^{-1}$ to support closed-form expressions. 
\subsubsection{$m = n$} the ($x,y$)-th element in $\textbf{U}_{kk'}^{m,m}$ can be given by \eqref{derivative_SE_mm} at the top of the next page 
\begin{figure*}[t!]
\begin{equation}
    \begin{array}{ll}
[\textbf{U}_{kk'}^{m,m}]_{xy}&\displaystyle=\left[\kappa_{m,r}
\mathbb{E}\Big{\{}
\hat{\textbf{G}}_{mk}^H{\textbf{G}}_{mk'}{\textbf{P}}_{k'}{\textbf{G}}_{mk'}^H\hat{\textbf{G}}_{mk}\Big{\}}\right]_{xy}=\kappa_{m,r}\mathbf{y}_{mk,p}^H(\mathbf{Z}_{mk}^{x:})^H{\textbf{G}}_{mk'}{\textbf{P}}_{k'}{\textbf{G}}_{mk'}^H\mathbf{Z}_{mk}^{y:}\mathbf{y}_{mk,p}
\\&\displaystyle=\underbrace{\kappa_{m,r}^2\tau_pp_p\kappa_{k',t}\sum\limits_{n=1}^{N_u}\xi_{k'n}\left[
\begin{array}{ll}
\bar{{\Delta}}_{mk'}^2\text{tr}\left(\bar{\textbf{K}}_{mkk',x}\textbf{A}_{mk'n}\right)\text{tr}\left(\textbf{A}_{mk'n}^H\bar{\textbf{K}}_{mkk',y}^H\right)+(\beta_{m}\beta_{k'})^2\text{tr}(\textbf{T}_{\omega_{k'}}^2)\text{tr}\left(\bar{\textbf{K}}_{mkk',x}\textbf{A}_{mk'n}\textbf{A}_{mk'n}^H\bar{\textbf{K}}_{mkk',y}^H\right)
\end{array}
\right]}_{k'\in\mathcal{P}_k}\\&\displaystyle+\kappa_{m,r}^2\sum\limits_{k''\in\mathcal{P}_k}\tau_pp_p\kappa_{k'',t}\sum\limits_{n=1}^{N_u}\xi_{k'n}\left[
\begin{array}{ll}
\bar{{\Delta}}_{mk'}\bar{{\Delta}}_{mk''}\text{tr}\left(\bar{\textbf{K}}_{mkk'',x}\textbf{A}_{mk'n}\textbf{A}_{mk'n}^H\bar{\textbf{K}}_{mkk'',y}^H\right)+\beta_{m}^2\beta_{k'}\beta_{k''}\text{tr}(\textbf{T}_{\omega_{k'}}\textbf{T}_{\omega_{k''}})\text{tr}\left(\textbf{X}_{mkk''mk'xn}\textbf{X}_{mkk''mk'yn}^H\right)
\end{array}
\right]
\\&\displaystyle+\kappa_{m,r}^2\sum\limits_{n=1}^{N_u}\xi_{k'n}\left[\begin{array}{ll}
\bar{{\Delta}}_{mk'}^2\text{tr}\left(\bar{\mathbf{\Upsilon}}_{mkk'k'nx}\bar{\mathbf{\Upsilon}}_{mkk'k'ny}^H\big{(}\textbf{I}_{N_u}\otimes(1-\kappa_{k',t})p_p\textbf{P}_{k'}\big{)}\right)\\+(\beta_{m}\beta_{k'})^2\text{tr}(\textbf{T}_{\omega_{k'}}^2)\text{tr}\left(({\textbf{Z}}_{mk}^x)^H\textbf{A}_{mk'n}\textbf{A}_{mk'n}^H{\textbf{Z}}_{mk}^y\big{(}\textbf{I}_{N_u}\otimes\text{tr}((1-\kappa_{k',t})p_p\textbf{P}_{k'}\textbf{R}_{k',t})\textbf{R}_{m,r}\big{)}\right)
\end{array}
\right]
\\&\displaystyle+\kappa_{m,r}^2\sum\limits_{k''=1}^K\sum\limits_{n=1}^{N_u}\xi_{k'n}\left[\begin{array}{ll}
\bar{{\Delta}}_{mk'}\bar{{\Delta}}_{mk''}\text{tr}\left(({\textbf{Z}}_{mk}^x)^H\textbf{A}_{mk'n}\textbf{A}_{mk'n}^H{\textbf{Z}}_{mk}^y\big{(}\textbf{I}_{N_u}\otimes\text{tr}((1-\kappa_{k'',t})p_p\textbf{P}_{k''}\textbf{R}_{k'',t})\textbf{R}_{m,r}\big{)}\right)\\+\beta_{m}^2\beta_{k'}\beta_{k''}\text{tr}(\textbf{T}_{\omega_{k'}}\textbf{T}_{\omega_{k''}})\text{tr}\left(\bar{\mathbf{\Upsilon}}_{mkk'k''nx}\bar{\mathbf{\Upsilon}}_{mkk'k''ny}^H\big{(}\textbf{I}_{N_u}\otimes(1-\kappa_{k'',t})p_p\textbf{P}_{k''}\big{)}\right)
\end{array}
\right]
\\&\displaystyle+\kappa_{m,r}\bar{\mathbf{\Delta}}_{mk'}\text{tr}\left(\textbf{R}_{k',t}\textbf{P}_{k'}\right)\text{tr}\left(\left(\textbf{I}_{N_u}\otimes\textbf{C}_{m|\{\textbf{G}_{mk}\}}\right)(\textbf{Z}_{mk}^{x:})^H\textbf{R}_{m,r}\textbf{Z}_{mk}^{y:}
\right)+\kappa_{m,r}\bar{\mathbf{\Delta}}_{mk'}\text{tr}\left(\textbf{R}_{k',t}\textbf{P}_{k'}\right)
\text{tr}\left(\sigma^2(\textbf{Z}_{mk}^{x:})^H\textbf{R}_{m,r}\textbf{Z}_{mk}^{y:}
\right),
        \end{array}
        \label{derivative_SE_mm}
\end{equation}
\vspace{-6 pt}
\hrulefill
\end{figure*}
with the help of the following formulas
\begin{equation}
    \begin{array}{ll}
    \displaystyle\bar{\textbf{K}}_{mkk',x}=\left(\textbf{R}_{k',t}^{1/2}\textbf{P}_{k'}^{1/2} \otimes{\textbf{R}}_{m,r}^{1/2}
    \right)\left(\textbf{Z}_{mk}^{x:}
    \right)^H\in\mathbb{C}^{N_{ap}N_u\times N_{ap}},
        \end{array}
        \label{K_matrix}
\end{equation}
\begin{equation}
    \begin{array}{ll}
    \displaystyle\textbf{A}_{mkn}=\left(\textbf{R}_{k,t}^{1/2}(n,:) \otimes{\textbf{R}}_{m,r}^{1/2}
    \right)\in\mathbb{C}^{N_{ap}\times N_{ap}N_u}.
        \end{array}
        \label{derivative_SE_04}
\end{equation}
Moreover, the elements in $\textbf{X}_{mkk'mkxn} \in\mathcal{C}^{N_u\times N_u}$ and ${\boldsymbol{\Upsilon}}_{mkk'k''nx}\in\mathbb{C}^{N_{u}\times N_u}$ follow
\begin{equation}
    \begin{array}{ll}
    \displaystyle[\textbf{X}_{mkk'mkxn}]_{ab}&\displaystyle=\text{tr}\left( \bar{\textbf{K}}_{mkk',x}^{a,:}\textbf{A}_{mkn}^{:,b}
    \right)
\\&\displaystyle=\text{tr}\left(\left(\textbf{R}_{k,t}^{1/2}(n,b) \textbf{R}_{k',t}^{1/2}(a,:)\textbf{P}_{k'}^{1/2} \otimes{\textbf{R}}_{m,r}
    \right)\left(\textbf{Z}_{mk}^{x:}
    \right)^H
    \right),
        \end{array}
        \label{derivative_SE_05}
\end{equation}
\begin{equation}
    \begin{array}{ll}
    \displaystyle[{\boldsymbol{\Upsilon}}_{mkk'k''nx}]_{ab}=
\text{tr}\left(\textbf{A}_{mk''a}^H\left(\textbf{Z}_{mk}^{x,b}
    \right)^H\textbf{A}_{mk'n}
\right), 
        \end{array}
        \label{derivative_SE_06}
\end{equation}
\begin{equation}
    \begin{array}{ll}
    \displaystyle\bar{\boldsymbol{\Upsilon}}_{mkk'k''nx}=\text{vec}\left({\boldsymbol{\Upsilon}}_{mkk'k''nx}\right).
        \end{array}
        \label{derivative_SE_05}
\end{equation}
where $\textbf{Z}_{mk}^{x:}\in\mathbb{C}^{N_{ap}\times N_{ap}N_u}$ and $\textbf{Z}_{mk}^{x,b}\in\mathbb{C}^{N_{ap}\times N_{ap}}$ can be given by
\begin{equation}
    \begin{array}{ll}
   \textbf{Z}_{mk}^{x:}=\textbf{Z}_{mk}\big{(}(x-1)N_{ap}+1:xN_{ap},1:N_{ap}N_u\big{)},
        \end{array}
        \label{derivative_SE_6}
\end{equation}
\begin{equation}
    \begin{array}{ll}
   \textbf{Z}_{mk}^{x,b}\displaystyle=\textbf{Z}_{mk}\big{(}(x-1)N_{ap}+1:xN_{ap},(b-1)N_{ap}+1:bN_{ap}\big{)}.
        \end{array}
        \label{derivative_SE_6}
\end{equation}

\subsubsection{$m \neq m'$} similarly, the ($x,y$)-th element in $\textbf{U}_{kk'}^{m,m'}$ is given by \eqref{derivative_SE_mn} at the top of this page.
\begin{figure*}[t!]
\begin{equation}
    \begin{array}{ll}
[\textbf{U}_{kk'}^{m,m'}]_{xy}&\displaystyle=\left[\sqrt{\kappa_{m,r}\kappa_{m',r}}
\mathbb{E}\Big{\{}
\hat{\textbf{G}}_{mk}^H{\textbf{G}}_{mk'}{\textbf{P}}_{k'}{\textbf{G}}_{m'k'}^H\hat{\textbf{G}}_{m'k}\Big{\}}\right]_{xy}=\sqrt{\kappa_{m,r}\kappa_{m',r}}\mathbf{y}_{mk,p}^H(\mathbf{Z}_{mk}^{x:})^H{\textbf{G}}_{mk'}{\textbf{P}}_{k'}{\textbf{G}}_{m'k'}^H\mathbf{Z}_{m'k}^{y:}\mathbf{y}_{m'k,p}
\\&\displaystyle=\underbrace{\kappa_{m,r}\kappa_{m',r}\tau_pp_p\kappa_{k',t}\sum\limits_{n=1}^{N_u}\xi_{k'n}
\bar{{\Delta}}_{mk'}\bar{{\Delta}}_{m'k'}\text{tr}\left(\bar{\textbf{K}}_{mkk',x}\textbf{A}_{mk'n}\right)\text{tr}\left(\textbf{A}_{m'k'n}^H\bar{\textbf{K}}_{m'kk',y}^H\right)
}_{k'\in\mathcal{P}_k}\\&\displaystyle+\kappa_{m,r}\kappa_{m',r}\sum\limits_{k''\in\mathcal{P}_k}\tau_pp_p\kappa_{k'',t}\sum\limits_{n=1}^{N_u}\xi_{k'n}
\beta_{m}\beta_{m'}\beta_{k'}\beta_{k''}\text{tr}(\textbf{T}_{\omega_{k'}}\textbf{T}_{\omega_{k''}})\text{tr}\left(\textbf{X}_{mkk''mk'xn}\textbf{X}_{m'kk''m'k'yn}^H\right)

\\&\displaystyle+\kappa_{m,r}\kappa_{m',r}\sum\limits_{n=1}^{N_u}\xi_{k'n}
\bar{{\Delta}}_{mk'}\bar{{\Delta}}_{m'k'}\text{tr}\left(\bar{\mathbf{\Upsilon}}_{mkk'k'nx}\bar{\mathbf{\Upsilon}}_{m'kk'k'ny}^H\big{(}\textbf{I}_{N_u}\otimes(1-\kappa_{k',t})p_p\textbf{P}_{k'}\big{)}\right)
\\&\displaystyle+\kappa_{m,r}\kappa_{m',r}\sum\limits_{k''=1}^K\sum\limits_{n=1}^{N_u}\xi_{k'n}
\beta_{m}\beta_{m'}\beta_{k'}\beta_{k''}\text{tr}(\textbf{T}_{\omega_{k'}}\textbf{T}_{\omega_{k''}})\text{tr}\left(\bar{\mathbf{\Upsilon}}_{mkk'k''nx}\bar{\mathbf{\Upsilon}}_{m'kk'k''ny}^H\big{(}\textbf{I}_{N_u}\otimes(1-\kappa_{k'',t})p_p\textbf{P}_{k''}\big{)}\right).
        \end{array}
        \label{derivative_SE_mn}
\end{equation}
\vspace{-6 pt}
\hrulefill
\end{figure*}
\subsection{Compute Receiver Distortion}
We introduce $\bar{\mathbf{\Gamma}}_{k}$ as the closed-form expression of ${\mathbf{\Gamma}}_{k}$. Then, the $m$-th sub-matrix in $\bar{\mathbf{\Gamma}}_{k}$, $\bar{\mathbf{\Gamma}}_{k}^m \in \mathbb{C}^{N_u \times N_u}$, can be expressed as \eqref{Gamma_kk} at the top of the next page
\begin{figure*}
    
    \begin{equation}
\begin{array}{ll}
     \displaystyle \bar{\mathbf{\Gamma}}_{k}^m & \displaystyle=\mathbb{E}\left\{\hat{\textbf{G}}_{mk}^H\bar{\textbf{C}}_{m|\{\textbf{G}_{mk}\}}\hat{\textbf{G}}_{mk}\right\}\displaystyle=\mathbb{E}\left\{{\left[ {\begin{matrix}
{\hat{\textbf{g}}_{mk1}^H\bar{\textbf{C}}_{m|\{\textbf{G}_{mk}\}}\hat{\textbf{g}}_{mk1}}&  {\hat{\textbf{g}}_{mk1}^H\bar{\textbf{C}}_{m|\{\textbf{G}_{mk}\}}\hat{\textbf{g}}_{mk2}}&\cdots &{\hat{\textbf{g}}_{mk1}^H\bar{\textbf{C}}_{m|\{\textbf{G}_{mk}\}}\hat{\textbf{g}}_{mkN_u}}\\
{\hat{\textbf{g}}_{mk2}^H\bar{\textbf{C}}_{m|\{\textbf{G}_{mk}\}}\hat{\textbf{g}}_{mk1}}& {\hat{\textbf{g}}_{mk2}^H\bar{\textbf{C}}_{m|\{\textbf{G}_{mk}\}}\hat{\textbf{g}}_{mk2}} &  &\vdots\\
 \vdots &  &\ddots &  {\vdots} \\
{{\hat{\textbf{g}}_{mkN_u}^H\bar{\textbf{C}}_{m|\{\textbf{G}_{mk}\}}\hat{\textbf{g}}_{mk1}}}& {{\hat{\textbf{g}}_{mkN_u}^H\bar{\textbf{C}}_{m|\{\textbf{G}_{mk}\}}\hat{\textbf{g}}_{mk2}}} & \cdots  &{\hat{\textbf{g}}_{mkN_u}^H\bar{\textbf{C}}_{m|\{\textbf{G}_{mk}\}}\hat{\textbf{g}}_{mkN_u}}
\end{matrix}} \right]}\right\},
\end{array}\label{Gamma_kk}
   \end{equation}
   \hrulefill
   \end{figure*}
 with the $n,n'$-th element $[\bar{\mathbf{\Gamma}}_{k}^m]_{nn'}=\mathbb{E}\left\{{\hat{\textbf{g}}_{mkn}^H\bar{\textbf{C}}_{m|\{\textbf{G}_{mk}\}}\hat{\textbf{g}}_{mkn'}}\right\}=\text{tr}\left(\mathbf{\Psi}_{mk}\left(\textbf{Z}_{mk}^{n:}\right)^H\bar{\textbf{C}}_{m|\{\textbf{G}_{mk}\}}\textbf{Z}_{mk}^{n':}
 \right)=\text{tr}\left(\bar{\textbf{C}}_{m|\{\textbf{G}_{mk}\}}\hat{\boldsymbol{\Delta}}_{mk}^{n',n}
 \right)$.

\subsection{Compute Noise}

According to \eqref{D_k_CF1}-\eqref{D_k_CF2}, we can compute the closed-form expression of ${\mathbf{\Lambda}}_{k}$, $\bar{\mathbf{\Lambda}}_{k}\in\mathbb{C}^{MN_u\times MN_u}$, as
\begin{equation}
\begin{array}{ll}
     \displaystyle \bar{\mathbf{\Lambda}}_{k}=\text{blkdiag}\left(
     \frac{\bar{{\textbf{H}}}_{1kk}}{\sqrt{\kappa_{1,r}\kappa_{k,t}}},...,\frac{\bar{{\textbf{H}}}_{Mkk}}{\sqrt{\kappa_{M,r}\kappa_{k,t}}}
     
     \right).
\end{array}\label{Lambda_kk_CF}
   \end{equation} 
\end{appendices}


%




\ifCLASSOPTIONcaptionsoff
  \newpage
\fi



\bibliographystyle{IEEEtran}
\bibliography{IEEEabrv,ref}

\begin{thebibliography}{10}
\providecommand{\url}[1]{#1}
\csname url@samestyle\endcsname
\providecommand{\newblock}{\relax}
\providecommand{\bibinfo}[2]{#2}
\providecommand{\BIBentrySTDinterwordspacing}{\spaceskip=0pt\relax}
\providecommand{\BIBentryALTinterwordstretchfactor}{4}
\providecommand{\BIBentryALTinterwordspacing}{\spaceskip=\fontdimen2\font plus
\BIBentryALTinterwordstretchfactor\fontdimen3\font minus \fontdimen4\font\relax}
\providecommand{\BIBforeignlanguage}[2]{{%
\expandafter\ifx\csname l@#1\endcsname\relax
\typeout{** WARNING: IEEEtran.bst: No hyphenation pattern has been}%
\typeout{** loaded for the language `#1'. Using the pattern for}%
\typeout{** the default language instead.}%
\else
\language=\csname l@#1\endcsname
\fi
#2}}
\providecommand{\BIBdecl}{\relax}
\BIBdecl

\bibitem{9570143}
X.~Mu, Y.~Liu, L.~Guo, J.~Lin, and R.~Schober, ``Simultaneously transmitting and reflecting {(STAR) RIS} aided wireless communications,'' \emph{IEEE Trans. Wireless Commun.}, vol.~21, no.~5, pp. 3083--3098, 2022.

\bibitem{10167480}
E.~Shi, J.~Zhang, D.~W.~K. Ng, and B.~Ai, ``Uplink performance of {RIS}-aided cell-free massive {MIMO} system with electromagnetic interference,'' \emph{IEEE J. Sel. Areas Commun.}, vol.~41, no.~8, pp. 2431--2445, 2023.

\bibitem{8388873}
J.~Qian, C.~Masouros, and A.~Garcia-Rodriguez, ``Partial {CSI} acquisition for size-constrained massive {MIMO} systems with user mobility,'' \emph{IEEE Trans. Veh. Technol.}, vol.~67, no.~9, pp. 9016--9020, 2018.

\bibitem{10163977}
M.~Xie, X.~Yu, Y.~Rui, K.~Wang, X.~Dang, and J.~Zhang, ``Performance analysis for user-centric cell-free massive {MIMO} systems with hardware impairments and multi-antenna users,'' \emph{IEEE Trans. Wireless Commun.}, vol.~23, no.~2, pp. 1243--1259, 2024.

\bibitem{9665300}
T.~Van~Chien, H.~Q. Ngo, S.~Chatzinotas, M.~Di~Renzo, and B.~Ottersten, ``Reconfigurable intelligent surface-assisted cell-free massive {MIMO} systems over spatially-correlated channels,'' \emph{IEEE Trans. Wireless Commun.}, vol.~21, no.~7, pp. 5106--5128, 2022.

\bibitem{10201892}
Q.~Sun, X.~Ji, Z.~Wang, X.~Chen, Y.~Yang, J.~Zhang, and K.-K. Wong, ``Uplink performance of hardware-impaired cell-free massive {MIMO} with multi-antenna users and superimposed pilots,'' \emph{IEEE Trans. Commun.}, vol.~71, no.~11, pp. 6711--6726, 2023.

\bibitem{9737367}
Z.~Wang, J.~Zhang, B.~Ai, C.~Yuen, and M.~Debbah, ``Uplink performance of cell-free massive {MIMO} with multi-antenna users over jointly-correlated rayleigh fading channels,'' \emph{IEEE Trans. Wireless Commun.}, vol.~21, no.~9, pp. 7391--7406, 2022.

\bibitem{10058895}
X.~Ma, D.~Zhang, M.~Xiao, C.~Huang, and Z.~Chen, ``Cooperative beamforming for {RIS}-aided cell-free massive {MIMO} networks,'' \emph{IEEE Trans. Wireless Commun.}, vol.~22, no.~11, pp. 7243--7258, 2023.

\bibitem{7827017}
H.~Q. Ngo, A.~Ashikhmin, H.~Yang, E.~G. Larsson, and T.~L. Marzetta, ``Cell-free massive {MIMO} versus small cells,'' \emph{IEEE Trans. Wireless Commun.}, vol.~16, no.~3, pp. 1834--1850, 2017.

\bibitem{9416909}
J.~Zheng, J.~Zhang, E.~Bj{\"o}rnson, and B.~Ai, ``Impact of channel aging on cell-free massive {MIMO} over spatially correlated channels,'' \emph{IEEE Trans. Wireless Commun.}, vol.~20, no.~10, pp. 6451--6466, 2021.

\bibitem{qian2024performance}
\BIBentryALTinterwordspacing
J.~Qian, R.~Murch, and K.~B. Letaief, ``Performance analysis of {STAR-RIS}-assisted cell-free massive {MIMO} systems with electromagnetic interference and phase errors,'' 2024. [Online]. Available: \url{https://arxiv.org/abs/2411.14030}
\BIBentrySTDinterwordspacing

\bibitem{8845768}
E.~Bj{\"o}rnson and L.~Sanguinetti, ``Making cell-free massive {MIMO} competitive with {MMSE} processing and centralized implementation,'' \emph{IEEE Trans. Wireless Commun.}, vol.~19, no.~1, pp. 77--90, 2020.

\bibitem{10297571}
A.~Papazafeiropoulos, H.~Q. Ngo, P.~Kourtessis, and S.~Chatzinotas, ``{STAR-RIS} assisted cell-free massive {MIMO} system under spatially-correlated channels,'' \emph{IEEE Trans. Veh. Tech.}, pp. 1--16, 2023.

\bibitem{9322151}
M.~Bashar, K.~Cumanan, A.~G. Burr, P.~Xiao, and M.~Di~Renzo, ``On the performance of reconfigurable intelligent surface-aided cell-free massive {MIMO} uplink,'' in \emph{Proc. IEEE GLOBECOM}, 2020, pp. 1--6.

\bibitem{9326394}
Q.~Wu, S.~Zhang, B.~Zheng, C.~You, and R.~Zhang, ``Intelligent reflecting surface-aided wireless communications: A tutorial,'' \emph{IEEE Trans. Commun.}, vol.~69, no.~5, pp. 3313--3351, 2021.

\bibitem{10225319}
Y.~Zhang, W.~Xia, H.~Zhao, G.~Zheng, S.~Lambotharan, and L.~Yang, ``Performance analysis of {RIS}-assisted cell-free massive {MIMO} systems with transceiver hardware impairments,'' \emph{IEEE Trans. Commun.}, pp. 1--1, 2023.

\bibitem{10316600}
Y.~Song, S.~Xu, R.~Xu, and B.~Ai, ``Weighted sum-rate maximization for multi-{STAR-RIS}-assisted mmwave cell-free networks,'' \emph{IEEE Trans. Veh. Technol}, vol.~73, no.~4, pp. 5304--5320, 2024.

\bibitem{9437234}
J.~Xu, Y.~Liu, X.~Mu, and O.~A. Dobre, ``{STAR-RISs:} simultaneous transmitting and reflecting reconfigurable intelligent surfaces,'' \emph{IEEE Commun. Lett.}, vol.~25, no.~9, pp. 3134--3138, 2021.

\bibitem{9690478}
Y.~Liu, X.~Mu, J.~Xu, R.~Schober, Y.~Hao, H.~V. Poor, and L.~Hanzo, ``{STAR}: Simultaneous transmission and reflection for {360°} coverage by intelligent surfaces,'' \emph{IEEE Wireless Commun.}, vol.~28, no.~6, pp. 102--109, 2021.

\bibitem{10264149}
X.~Ma, X.~Lei, P.~T. Mathiopoulos, and D.~B. da~Costa, ``Active {STAR-RIS} aided cell-free massive {MIMO}: A performance study,'' \emph{IEEE Trans. Veh. Technol.}, pp. 1--6, 2023.

\bibitem{10418910}
Y.~Lu, J.~Zhang, J.~Zheng, H.~Xiao, and B.~Ai, ``Performance analysis of {RIS}-assisted communications with hardware impairments and channel aging,'' \emph{IEEE Trans. Commun.}, vol.~72, no.~6, pp. 3720--3735, 2024.

\bibitem{8476516}
J.~Zhang, Y.~Wei, E.~Bj{\"o}rnson, Y.~Han, and S.~Jin, ``Performance analysis and power control of cell-free massive {MIMO} systems with hardware impairments,'' \emph{IEEE Access}, vol.~6, pp. 55\,302--55\,314, 2018.

\bibitem{9528977}
A.~Papazafeiropoulos, E.~Bj{\"o}rnson, P.~Kourtessis, S.~Chatzinotas, and J.~M. Senior, ``Scalable cell-free massive {MIMO} systems: Impact of hardware impairments,'' \emph{IEEE Trans. Veh. Tech.}, vol.~70, no.~10, pp. 9701--9715, 2021.

\bibitem{10445267}
X.~Chen, C.~Zhou, Z.~Wang, T.~Zhao, Q.~Sun, C.~Xu, and J.~Zhang, ``Energy efficiency of wireless-powered cell-free {mMIMO} with hardware impairments,'' \emph{IEEE Trans. Commun.}, vol.~72, no.~7, pp. 4446--4458, 2024.

\bibitem{10475146}
X.~Luo, Z.~Jiang, F.~Xu, X.~Li, G.~Zhu, and K.~Shen, ``Sum-rate maximization for {STAR-RIS}-assisted multi-user networks with hardware impairments,'' \emph{IEEE Wireless Commun. Lett.}, vol.~13, no.~5, pp. 1503--1507, 2024.

\bibitem{10264820}
A.~Papazafeiropoulos, P.~Kourtessis, and S.~Chatzinotas, ``{Max-Min} sinr analysis of {STAR-RIS} assisted massive {MIMO} systems with hardware impairments,'' \emph{IEEE Trans. Wireless Commun.}, vol.~23, no.~5, pp. 4255--4268, 2024.

\bibitem{sui2024starrisaidedcellfreemassivemimo}
\BIBentryALTinterwordspacing
Z.~Sui, H.~Q. Ngo, and M.~Matthaiou, ``{STAR-RIS}-aided cell-free massive {MIMO} with imperfect hardware,'' 2024. [Online]. Available: \url{https://arxiv.org/abs/2408.14436}
\BIBentrySTDinterwordspacing

\bibitem{10841966}
Z.~Sui, H.~Q. Ngo, M.~Matthaiou, and L.~Hanzo, ``Performance analysis and optimization of {STAR-RIS}-aided cell-free massive {MIMO} systems relying on imperfect hardware,'' \emph{IEEE Trans. Wireless Commun.}, pp. 1--1, 2025.

\bibitem{10326460}
J.~Dai, J.~Ge, K.~Zhi, C.~Pan, Z.~Zhang, J.~Wang, and X.~You, ``Two-timescale transmission design for {RIS}-aided cell-free massive {MIMO} systems,'' \emph{IEEE Trans. Wireless Commun.}, vol.~23, no.~6, pp. 6498--6517, 2024.

\bibitem{10571171}
J.~Qian, C.~Zhang, K.~B. Letaief, and R.~Murch, ``The effect of spatial correlation and mutual coupling on cell-free massive {MIMO},'' in \emph{Proc. IEEE WCNC}, 2024, pp. 01--06.

\bibitem{9079911}
T.~C. Mai, H.~Q. Ngo, and T.~Q. Duong, ``Downlink spectral efficiency of cell-free massive {MIMO} systems with multi-antenna users,'' \emph{IEEE Trans. Commun.}, vol.~68, no.~8, pp. 4803--4815, 2020.

\bibitem{10620555}
G.~Dogim, E.~A. Maher, and A.~El-Mahdy, ``Multi-{RIS} aided cell-free {MIMO} networks with imperfect {CSI},'' in \emph{Proc. IEEE ITC-Egypt}, 2024, pp. 519--524.

\bibitem{9905943}
Y.~Zhang, J.~Zhang, H.~Xiao, D.~W.~K. Ng, and B.~Ai, ``Channel aging-aware precoding for {RIS}-aided multi-user communications,'' \emph{IEEE Trans. Veh. Technol.}, vol.~72, no.~2, pp. 1997--2008, 2023.

\bibitem{9875036}
A.~Papazafeiropoulos, I.~Krikidis, and P.~Kourtessis, ``Impact of channel aging on reconfigurable intelligent surface aided massive {MIMO} systems with statistical {CSI},'' \emph{IEEE Trans. Veh. Technol.}, vol.~72, no.~1, pp. 689--703, 2023.

\bibitem{7500452}
X.~Li, E.~Bj{\"o}rnson, S.~Zhou, and J.~Wang, ``Massive {MIMO} with multi-antenna users: When are additional user antennas beneficial?'' in \emph{Proc. IEEE ICT}, 2016, pp. 1--6.

\bibitem{9685245}
{\"O}.~T. Demir and E.~Bj{\"o}rnson, ``{RIS}-assisted massive {MIMO} with multi-specular spatially correlated fading,'' in \emph{Proc. IEEE GLOBECOM}, 2021, pp. 1--6.

\bibitem{9598875}
A.~de~Jesus~Torres, L.~Sanguinetti, and E.~Bj{\"o}rnson, ``Electromagnetic interference in {RIS}-aided communications,'' \emph{IEEE Wireless Commun. Lett.}, vol.~11, no.~4, pp. 668--672, 2022.

\bibitem{9300189}
E.~Bj{\"o}rnson and L.~Sanguinetti, ``Rayleigh fading modeling and channel hardening for reconfigurable intelligent surfaces,'' \emph{IEEE Wirel. Commun. Lett.}, vol.~10, no.~4, pp. 830--834, 2021.

\bibitem{8696221}
T.~C. Mai, H.~Q. Ngo, and T.~Q. Duong, ``Uplink spectral efficiency of cell-free massive {MIMO} with multi-antenna users,'' in \emph{Proc. IEEE SigTelCom}, 2019, pp. 126--129.

\bibitem{8891922}
H.~Masoumi and M.~J. Emadi, ``Performance analysis of cell-free massive {MIMO} system with limited fronthaul capacity and hardware impairments,'' \emph{IEEE Trans. Wireless Commun.}, vol.~19, no.~2, pp. 1038--1053, 2020.

\bibitem{9459571}
Y.~Zhang, M.~Zhou, H.~Zhao, L.~Yang, and H.~Zhu, ``Spectral efficiency of superimposed pilots in cell-free massive {MIMO} systems with hardware impairments,'' \emph{China Communications}, vol.~18, no.~6, pp. 146--161, 2021.

\bibitem{10001172}
M.~Xie, X.~Yu, J.~Xu, and X.~Dang, ``Low-complexity channel estimation scheme for cell-free massive {MIMO} with hardware impairment,'' in \emph{Proc. IEEE GLOBECOM}, 2022, pp. 711--716.

\bibitem{8171057}
J.~Zhang, Y.~Wei, E.~Bj{\"o}rnson, Y.~Han, and X.~Li, ``Spectral and energy efficiency of cell-free massive {MIMO} systems with hardware impairments,'' in \emph{Proc. IEEE WCSP}, 2017, pp. 1--6.

\bibitem{951380}
S.~Loyka, ``Channel capacity of {MIMO} architecture using the exponential correlation matrix,'' \emph{IEEE Commun. Lett.}, vol.~5, no.~9, pp. 369--371, 2001.

\bibitem{10449720}
Y.~Chen, W.~Xia, J.~Zhang, and Y.~Zhu, ``Joint learning of channel estimation and beamforming for cell-free massive mimo systems,'' \emph{IEEE Wireless Commun. Lett.}, vol.~13, no.~5, pp. 1359--1363, 2024.

\end{thebibliography}
\end{document}